\documentclass[useAMS,usenatbib]{mn2e}

\usepackage{amssymb}
\usepackage{amsmath}
\usepackage{graphicx}
\usepackage{textcomp}
\usepackage{natbib}
\usepackage{color}
\def\Cmat{{\mathbf C}}

\def\Ymat{{\mathbf Y}}

\title[Errors in redshift-space distortion measurements]{Statistical
  and systematic errors in redshift-space distortion measurements from large surveys}
\author[D. Bianchi, et al.]{D. Bianchi$^{1,2}$\thanks{E-mail:
davide.bianchi@brera.inaf.it}, 
L. Guzzo$^{2}$,
E. Branchini$^{2,3,4}$,
E. Majerotto$^{2,5,6}$,
S. de la Torre$^{7}$,
\newauthor
F. Marulli$^{8,9,10}$,
L. Moscardini$^{8,9,10}$,
and R. E. Angulo$^{11}$\\
$^{1}$Dipartimento di Fisica, Universit\`a degli Studi di Milano, via Celoria 16, I-20133 Milano, Italy\\
$^{2}$INAF - Osservatorio Astronomico di Brera, Via Bianchi 46, I-23807 Merate (LC), Italy\\
$^{3}$Dipartimento di Fisica, Universit\`a degli Studi ``Roma Tre'', via della Vasca Navale 84, I-00146 Roma, Italy\\
$^{4}$INFN, Sezione di Roma Tre, via della Vasca Navale 84, I-00146, Roma, Italy\\
$^{5}$Instituto de F{\'i}sica Te{\'o}rica (UAM/CSIC), Universidad Aut{\'o}noma de Madrid, Cantoblanco, 28049 Madrid, Spain\\
$^{6}$Departamento de F{\'i}sica Te{\'o}rica (UAM), Universidad Aut{\'o}noma de Madrid, Cantoblanco, 28049 Madrid, Spain\\
$^{7}$SUPA\thanks{Scottish Universities Physics Alliance}, Institute for Astronomy, University of Edinburgh, Royal Observatory, Blackford Hill, EH9 3HJ Edinburgh, UK\\
$^{8}$Dipartimento di Astronomia, Alma Mater Studiorum-Universit\`a di Bologna, via Ranzani 1, I-40127 Bologna, Italy\\
$^{9}$INAF - Osservatorio Astronomico di Bologna, via Ranzani 1, I-40127 Bologna, Italy\\
$^{10}$INFN, Sezione di Bologna, viale Berti Pichat 6/2, I-40127 Bologna, Italy\\
$^{11}$Max-Planck Institut f\"ur Astrophysics, D-85748, Garching b. M\"unchen, Germany\\
}

\begin{document}

\date{Accepted 2012 September 9. Received 2012 September 9; in original form 2012 March 8}

\pagerange{\pageref{firstpage}--\pageref{lastpage}} \pubyear{2012}

\maketitle

\label{firstpage}
\begin{abstract}
We investigate the impact of statistical and systematic errors on measurements of linear
redshift-space distortions (RSD) in future cosmological surveys by analysing
large catalogues of dark-matter halos from the BASICC simulation.
These allow us to estimate the dependence of errors on typical survey
properties, as volume, galaxy density and mass (i.e.  bias factor) of
the adopted tracer.
We find that measures of the specific growth rate
$\beta=f/b$ using the Hamilton/Kaiser
harmonic expansion of the redshift-space correlation function
$\xi(r_p,\pi)$ on scales larger than $3 \ h^{-1} \ \text{Mpc}$ are typically
under-estimated  by up to 10\% for galaxy sized halos. This is 
significantly larger than the corresponding statistical errors, which
amount to a few percent, indicating the importance of non-linear
improvements to the Kaiser model, to obtain accurate measurements 
of the growth rate. The systematic error shows a
diminishing trend with increasing bias value (i.e. mass) of the halos
considered. 
We compare the amplitude and trends of statistical errors as a
function of survey 
parameters to predictions obtained with the Fisher information matrix
technique. This is what is usually adopted to produce RSD forecasts, based on the FKP
prescription for the errors on the power spectrum.  We show that this
produces parameter errors fairly similar to the standard deviations
from the halo catalogues, provided it is applied to strictly linear
scales in Fourier space ($k<0.2 \ h \ \text{Mpc}^{-1}$).
Finally, we combine our measurements to define and calibrate an
accurate scaling formula for the relative error on $\beta$ as a
function of the same parameters, which closely matches the simulation
results in all explored regimes. This provides a handy and plausibly
more realistic alternative to the Fisher matrix approach, to quickly and accurately
predict statistical errors on RSD expected from future surveys.
\end{abstract} 


\section{Introduction}
Galaxy clustering as measured in redshift-space contains the imprint
of the linear growth rate of structure $f(z)$, in the form of a
measurable large-scale anisotropy \citep{1987MNRAS.227....1K}.  This
is produced by the coherent peculiar velocity flows towards
overdensities, which add an angle-dependent contribution to the measured redshift.
In linear theory, these {\it redshift-space distortions} (RSD) in the
clustering pattern can be quantified in terms of the ratio
$\beta(z)=f(z)/b(z)$ (where $b$ is the linear bias of the sample of galaxies considered).
A value for $\beta$ can be obtained by modeling the anisotropy of the redshift-space
two-point correlation function $\xi(r_p , \pi)$ (where $r_p$ and $\pi$ are the separations perpendicular and parallel
to the line of sight) or, equivalently, of the power spectrum (see \citet{Ha} for a review).
Since $b$ can be defined as the ratio of the {\it rms} galaxy clustering
amplitude to that of the underlying matter, $b\approx
\sigma_8^{gal}/\sigma_8^{mass}$, the measured product $\beta \times
\sigma_8^{gal}$ is equivalent to the predicted combination $f(z)\times
\sigma_8^{mass}(z)$ \citep{SP}.  The latter is a prediction depending
on the gravity theory, once normalized to the amplitude of matter
fluctuations at the given epoch, e.g. using CMB measurements. 

Measurements of the growth rate $f(z)$ are
crucial to pinpoint the origin of cosmic acceleration, distinguishing
whether it requires the addition of ``dark energy'' in the cosmic budget, or
rather a modification of General Relativity.  These two
radically alternative scenarios are degenerate when considering the
expansion rate $H(z)$ alone, as yielded, e.g., by the Hubble diagram
of Type Ia supernova (e.g. \citealt{Ri, Pe}) or Baryonic Acoustic
Oscillations (BAO, e.g \citealt{2010MNRAS.401.2148P}).  Although the
RSD effect is well known since long, its important
potential in the context of dark energy studies has been fully
appreciated only recently \citep{ZhangNEW, 2008Natur.451..541G}. This led to
a true renaissance of interest in this technique
\citep{Wn,Li2,Ne,Ac,SP,2009MNRAS.397.1348W,PW,2009MNRAS.393.1183C,2011MNRAS.415.2876B}, such that 
RSD have quickly become one of the most promising probes for future large
dark energy surveys. This is the case of the recently approved ESA Euclid mission 
\citep{2011arXiv1110.3193L}, which is expected
to reach statistical errors of a few percent on measurements of
$f(z)$ in several redshift bins out to $z=2$ using this technique
(coupled to similar precisions with the complementary weak-lensing experiment). 

In general, forecasts of
the statistical precision reachable by future projects on the 
measurements of different cosmological parameters have been produced
through widespread application of the so-called Fisher information
matrix technique \citep{1997PhRvL..79.3806T}.  This has also been done
specifically for RSD estimates of the growth rate and related
quantities \citep{Wn,Li2,2009MNRAS.397.1348W,PW,2009JCAP...10..007M}.  One limitation of these
forecasts is that they necessarily imply some idealized assumptions
(e.g. on the Gaussian nature of errors) and have not been verified, in 
general, against systematic numerical tests.  This is not easily
doable in general, given the large size of planned surveys. 
A first attempt to produce general forecasts based on numerical
experiments was presented by \citet{2008Natur.451..541G}, who used
mock surveys built from the Millennium simulation \citep{springel05} to numerically
estimate the random and systematic errors affecting their measurement
of the growth rate from the VIMOS VLT Deep Survey.  Using a grid of reference survey
configurations, they calibrated an approximated scaling relation for
the relative error on $\beta$ as a function of survey volume and mean
density.  The range of parameters explored in this case was however
limited, and one specific class of galaxies only (i.e. bias) was
analyzed.  

The second crucial aspect to be taken into consideration when 
evaluating Fisher matrix predictions, is that they only consider statistical
errors and cannot say anything about the importance of systematic
effects, i.e.  on the {\it accuracy} of the expected estimates.  This
is clearly a key issue for projects aiming at percent or sub-percent
precisions, for which systematic errors will be the dominant source of
uncertainty. 

In fact, a number of works in recent years suggest that the standard
linear Kaiser description of RSD is not sufficiently accurate on 
quasi-linear scales ($\approx 5-50 \ h^{-1} \text{Mpc}$) where it is
routinely applied
(\citealt{2004PhRvD..70h3007S, 2006MNRAS.368...85T, 2010arXiv1006.0699T,
  2011MNRAS.410.2081J}).   
Various non-linear corrections are proposed in these papers, 
the difficulty often being their practical implementation in the
analysis of real data, in particular in configuration space \citep{2012arXiv1202.5559D}. 
One may hope that in the future, with surveys covering much larger
volumes, it will be possible to limit the analysis to very large
scales, where the simple linear description should be
adequate. Still, ongoing surveys like Wigglez
\citep{2011MNRAS.415.2876B}, BOSS \citep{2011AJ....142...72E} and
VIPERS (Guzzo et al., in preparation), will still need to rely on the
clustering signal at intermediate scales to model RSD. 

Here, we shall address in a more systematic and extended
way the impact of random and systematic errors on growth rate
measurements using RSD in future surveys. We shall compare the results
directly to Fisher matrix predictions, thoroughly exploring the
dependence of statistical errors on the survey 
parameters, including also, in addition to volume and density, the
bias parameter of the galaxies used.  This is also relevant, as one
could wonder which kind of objects would be best suited to measure RSD
in a future project.  These will include using halos of different mass
(i.e. bias), up to those traced by groups and clusters of galaxies.
Potentially, using groups and clusters to measure RSD could be
particularly interesting in view of massive galaxy redshift surveys as
that expected from Euclid \citep{2011arXiv1110.3193L}, which can be used to build large
catalogues of optically-selected clusters with measured redshifts.
A similar opportunity will be offered by future X-ray surveys,
such as those expected from the E-Rosita mission
\citep{2011MSAIS..17..159C}, although in that case, mean 
cluster redshifts will have to be measured first.  

This paper is complementary to the parallel work of
\citet{marulli2012}, where we investigate the impact on RDS of
redshift errors and explore how to disentangle geometrical distortions
introduced by the uncertainty of the underlying geometry of the
Universe -- i.e. the Alcock-Paczynski effect
\citep{1979Natur.281..358A} --  on measurements of RSD.
Also, while we were completing our work, independent important 
contributions in the same direction appeared in the literature by
\citet{2011ApJ...726....5O} and \citet{2011arXiv1105.1194K}. 

The paper is organized as follows.
In \S~\ref{sec sims} we describe the simulations used and the mass-selected subsamples we defined; in
\S~\ref{sec rsd} we discuss the technical tools used to estimate and model the
two-point correlation function in redshift space, $\xi(r_p , \pi)$, and
to estimate the intrinsic values of bias and distortion to be used as
reference; in \S~\ref{sec err} we present the measured $\xi(r_p , \pi)$
and show the resulting statistical and systematic errors on $\beta$, as a
function of the halo bias; here we discuss in detail how  well
objects related to high-bias halos, as groups and clusters, can be
used to measure RSD; in \S~\ref{sec form} we organise all our results into
a compact analytic formula as a function of galaxy density, bias
and survey volume; we then directly compare these
results to the predictions of a Fisher matrix code; finally we summarize our
results in \S~\ref{sec concl}.

\section{Simulated data and Error estimation}\label{sec sims}

\subsection{Halo catalogues from the BASICC simulations \label{sec BASICC}} 
The core of this study is based on the
high-resolution Baryonic Acoustic-oscillation Simulations at the
Institute for Computational Cosmology (BASICC) of
\citet{2008MNRAS.383..755A}, which used $1448^3$ particles of mass
$5.49 \times 10^{10} \, h^{-1} \, M_\odot$ to follow the growth of
structure in dark matter in a periodic box of side $1340 \, h^{-1}
\text{Mpc}$.  The simulation volume was chosen to allow for growth of
fluctuations to be modelled accurately on a wide range of scales
including those of BAO.  The very large volume of the box also
allows us to extract accurate measurements of the clustering of massive
halos.  The mass resolution of the simulation is high enough to
resolve halos that should host the galaxies expected to be seen
in forthcoming high-redhift galaxy surveys (as e.g. Luminous Red
Galaxies in the case of SDSS-III BOSS).  The cosmological parameters adopted
are broadly consistent with recent data from the cosmic microwave
background and the power spectrum of galaxy clustering
(\citealt{2006MNRAS.366..189S}): the matter density parameter is
$\Omega_M = 0.25$, the cosmological constant density parameter
$\Omega_\Lambda = 0.75$, the normalization of density fluctuations,
expressed in terms of their linear amplitude in spheres of radius $8
\, h^{-1} \text{Mpc}$ at the present day $\sigma_8 = 0.9$, the
primordial spectral index $n_s = 1$, the dark energy equation of state
$w = -1$, and the reduced Hubble constant $h = H_0 / (100 \, \text{km} \,
\text{s}^{-1} \, \text{Mpc}^{-1}) = 0.73$.  We note the high value of
normalization of the power spectrum $\sigma_8$, with respect to more
recent WMAP estimates ($\sigma_8=0.801 \pm 0.030$, \citealt{2011ApJS..192...16L}). This has no effect on the results discussed
here (but see \citet{angulo_white 2010} for a method to scale
self-consistently the output of a simulation to a different background cosmology).
Outputs of the particle positions and velocities are stored from the simulations at selected redshifts. 
Dark matter halos are identified using a Friends-of-Friends (FOF) percolation algorithm \citep{1985ApJ...292..371D} with a linking length of $0.2$ times the mean particle separation.
Position and velocity are given by the values of the center of mass.
In this paper, only groups with al least $N_{part} = 20$ particles are considered
(i.e only halos with mass $M_{halo} \ge 1.10 \times 10^{12} \ h^{-1} \ M_\odot$).
This limit provides reliable samples in term of their abundance and clustering, which we
checked by comparing the halo mass function and correlation function
against \citet{2001MNRAS.321..372J} and \citet{2010ApJ...724..878T} respectively.

We use the complete catalogue of halos of the simulation at $z=1$,
from which we select sub-samples with different mass thresholds (i.e.
number of particles). This corresponds to samples with different bias values.  
Table \ref{tab halo masses} reports the main features of these catalogues.
\begin{table}
  \begin{center}
    \begin{tabular}{cccc}
      $N_{cut}$ & $M_{cut} \ [h^{-1} \ M_\odot$] & $\mathcal{N}_{tot}$ & $n \ [h^3 \ \text{Mpc}^{-3}]$\\
      \hline
      $20$ & $1.10 \times 10^{12}$ & $7483318$ & $3.11 \times 10^{-3}$\\
      $30$ & $1.65 \times 10^{12}$ & $4897539$ & $2.04 \times 10^{-3}$\\
      $45$ & $2.47 \times 10^{12}$ & $3158088$ & $1.31 \times 10^{-3}$\\
      $63$ & $3.46 \times 10^{12}$ & $2164960$ & $9.00 \times 10^{-4}$\\
      $91$ & $5.00 \times 10^{12}$ & $1411957$ & $5.87 \times 10^{-4}$\\
      $136$ & $7.47 \times 10^{12}$ & $866034$ & $3.60 \times 10^{-4}$\\
      $182$ & $9.99 \times 10^{12}$ & $597371$ & $2.48 \times 10^{-4}$\\
      $236$ & $1.30 \times 10^{13}$ & $423511$ & $1.76 \times 10^{-4}$\\
      $310$ & $1.70 \times 10^{13}$ & $290155$ & $1.21 \times 10^{-4}$\\
      $364$ & $2.00 \times 10^{13}$ & $230401$ & $9.58 \times 10^{-5}$\\
      $455$ & $2.50 \times 10^{13}$ & $165267$ & $6.87 \times 10^{-5}$\\
      $546$ & $3.00 \times 10^{13}$ & $124497$ & $5.17 \times 10^{-5}$\\
    \end{tabular}
  \end{center}
  \caption{Properties of the halo catalogues used in the analysis.
  $N_{cut}$ is the threshold value of $N_{part}$, e.g. the catalogue
  $N_{cut} = 20$ is the set of groups (i.e. halos) with at least $20$
  DM particles; $M_{cut}$ is the corresponding threshold mass;
  $\mathcal{N}_{tot}$ is the total number of halos (i.e. the number
  of halos with $M_{halo} \ge M_{cut}$); $n$ is the number density
  (i.e. $n = \mathcal{N}_{tot} / V$, where $V = 1340^3 \ h^{-3}
  \text{Mpc}^3$ is the simulation volume).} 
 \label{tab halo masses}
\end{table}
In the following we shall refer to a given catalogue by its threshold
mass $M_{cut}$ (i.e. the mass of the least massive halo belonging to
that catalogue). 
We also use the complete dark matter sample
(hereafter DM), including more than $3 \times 10^{9}$
particles\footnote{Such a number of points involves very long
  computational times when calculating, e.g., a two-point correlation
  function. To overcome this problem, we often use a sparsely sampled
  sub-set of the DM catalogue. In order to limit the impact of shot-noise,
  we nevertheless always keep the DM samples denser than the least dense halo
  catalogue (i.e. $M_{cut} = 1.10 \times 10^{12} \ h^{-1} \ M_\odot$).
  We verified directly on a subset that our results do not effectively depend on the level of DM dilution.}.
For each catalogue, we split the whole (cubical) box of
the simulation into $N_{split}^3$ sub-cubes ($N_{split}=3$ unless
otherwise stated). Each sub-cube ideally represents a different
realization of the same portion of the Universe, so that we are able
to estimate the expected precision on a quantity of cosmological
interest through its scatter among the sub-cubes.  Using $N_{split}=3$
is a compromise between having a better statistics from a larger
number of sub-samples (at the price of not sampling some very large
scales), and covering even larger scales (with $N_{split}=2$), but with
fewer statistics.
In general, there are large-scale modes shared between the sub-cubes.
As a consequence, our assumption that each sub-sample can be treated as an independent realization breaks down on such scales.
To overcome this problem, we limit our analysis to scales much smaller than the size of the sub-cubes.

This analysis concentrates at $z=1$, because this is central to the
range of redshifts that will become more and more explored by surveys
of the next generation. This includes galaxies, but also surveys of
clusters of galaxies, as those that should be possible with the
eRosita satellite, possibly due to launch in 2013.  Exploring the expectations
from RSD studies using high-bias objects, corresponding e.g. to groups of
galaxies, is one of the main themes of this paper.

\subsection{Simulating redshift-space observations\label{sec redshift-space}}
For our measurements we need to simulate redshift-space observations.
In other words, we have to \textquotedblleft observe\textquotedblright
\ the simulations as if the only information about the distance of an
object was given by its redshift.  For this purpose we center the
sample (i.e. one of the sub-cubes) at a distance given by 
\begin{eqnarray}
  \label{eq distance}
  D_1&=& D(z=1) = \int_0^{z=1} \frac{c}{H(z')}dz' \nonumber\\
  &=& \int_0^{z=1} \frac{c}{H_0 \sqrt{\Omega_M + \Omega_\Lambda {(1 + z')}^3}}dz' \; ,
\end{eqnarray}
where the last equality holds for the flat $\Lambda$CDM cosmology of
the simulation.
More explicitly, we transform the positions $(X_i,Y_i,Z_i)$ of an object in a
sub-cube of side $L$, into new comoving coordinates 
\begin{eqnarray}
  -\dfrac{L}{2} \le & X_i & \le \dfrac{L}{2} \; , \nonumber \\
  D_1 - \dfrac{L}{2} \le & Y_i & \le D_1 + \dfrac{L}{2} \; , \\
  -\dfrac{L}{2} \le & Z_i & \le \dfrac{L}{2} \nonumber \; ,
\end{eqnarray}
where we arbitrarily choose the direction of the $Y$ axis for the translation
($Z$ represents a coordinate, not to be confused with the redshift
$z$). 
This procedure assigns to each object a comoving distance in real space
$D_i=\sqrt{X_i^2+Y_i^2+Z_i^2}$, 
hence, inverting Eq. (\ref{eq distance}), a cosmological (undistorted) 
redshift $z_i$. We then add the Doppler contribution
to obtain the ``observed'' redshift, as
\begin{equation}
  \label{eq z_dist}
  \hat{z_i} = z_i + \frac{v_r}{c} (1 + z_i) \; ,
\end{equation}
where $v_r$ is the line-of-sight peculiar velocity.
Using $\hat{z_i}$ instead of $z_i$ to compute the comoving distance of
an object gives its redshift-space coordinate.  
Finally, in order to eliminate the blurring effect introduced at the
borders of the cube, we trim a slice of $10 \ h^{-1} \text{Mpc}$ from
all sides, a value about three times larger than typical pairwise
velocity dispersion.

\section[]{Measuring Redshift-Space Distortions} \label{sec rsd}

\subsection{Modelling linear and non-linear distortions}
In a fundamental paper, \citet{1987MNRAS.227....1K} showed that, in the
linear regime, the redshift-space modification of the observed clustering
pattern due to coherent infall velocities takes a simple form in
Fourier space: 
\begin{equation}
 \label{eq Kaiser spectrum beta}
 P_S(k,\mu_k) = {(1 + \beta \mu_k^2)}^2 P_R(k) \; ,
\end{equation}
where $P$ is the power spectrum (subscripts $R$ and $S$ denote
respectively quantities in real and redshift space), $\mu_k$ is the
cosine of the angle between the line of sight and the wave vector
$\vec{k}$ and $\beta = f / b$ is the distortion factor, where $f =
\frac{d \log G}{d \log a}$ and $G$ is the linear growth factor of 
density perturbations.
\citet{1992ApJ...385L...5H} translated this result into configuration
space (i.e. in terms of correlation function, $\xi$): 
\begin{equation}
 \xi_S^{(L)}(r_p,\pi) = \xi_0(r) \mathcal{P}_0(\mu) + \xi_2(r) \mathcal{P}_2(\mu) + \xi_4(r) \mathcal{P}_4(\mu) \; ,
 \label{eq lin model}
\end{equation}
where $r_p$ and $\pi$ are the separations perpendicular and parallel
to the line of sight, $\mu$ is the cosine of the angle between the
separation vector and the line of sight $\mu = \cos\theta = \pi / r$,
$\mathcal{P}_i$ are Legendre Polynomials and $\xi_i$ are the multipole moments of
$\xi(r_p,\pi)$, which can be expressed as  
\begin{eqnarray}
 \xi_0(r) & = & \bigg(1 + \frac{2}{3}\beta + \frac{1}{5}\beta^2 \bigg) \xi(r) \\
 \xi_2(r) & = & \bigg(\frac{4}{3}\beta + \frac{4}{7}\beta^2 \bigg) [\xi(r) - \bar{\xi}(r)] \\
 \xi_4(r) & = & \frac{8}{35}\beta^2 \bigg[\xi(r) + \frac{5}{2}\bar{\xi}(r) - \frac{7}{2}\bar{\bar{\xi}}(r) \bigg] \; ,
\end{eqnarray}
where
\begin{eqnarray}
 \label{eq xi bar}
 \bar{\xi} & = & \frac{3}{r^3} \int_0^r \xi(t) t^2 dt \\
 \label{eq xi barbar}
 \bar{\bar{\xi}} & = & \frac{5}{r^5} \int_0^r \xi(t) t^4 dt \; .
\end{eqnarray}
The superscript $L$ reminds us that Eq. (\ref{eq lin model}) holds
only in linear regime. 
A full model, accounting for both linear and non-linear motions, is
obtained empirically, through a convolution with the distribution
function of random pairwise velocities along the line of sight
$\varphi(v)$: 
\begin{equation}
 \label{eq lin-exp model}
 \xi_S(r_p,\pi) = \int_{-\infty}^{+\infty} \xi_S^{(L)}\biggl[r_p,\pi-\dfrac{v(1+z)}{H(z)}\biggr] \varphi(v) dv \; ,
\end{equation}
where $z$ is the redshift and $H(z)$ is the Hubble function
\citep{1983ApJ...267..465D, 1994MNRAS.267..927F,Peacock}.  
We represent $\varphi(v)$ by an exponential form, consistent with 
observations and N-body simulations
(e.g. \citealt{1994ApJ...431..559Z}),  
\begin{equation}
 \label{eq vel distribution}
 \varphi(v) = \dfrac{1}{\sigma_{12} \sqrt{2}} e^{- \dfrac{\sqrt{2} |v|}{\sigma_{12}}} \; ,
\end{equation}
where $\sigma_{12}$ is a pairwise velocity dipersion.  We note in
passing that the use of a Gaussian form for $\varphi(v)$ is in some
cases to be preferred, as e.g. when large redshift measurement errors affects
the catalogues to be analyzed. This is discussed in detail in \citet{marulli2012}
Hereafter we shall refer to Eq. (\ref{eq lin model}) and Eq. (\ref{eq
  lin-exp model}) as the linear and linear-exponential model, respectively. 
Moreover, in order to simplify the notations, we shall refer to the
real- and redshift-space correlation functions just as $\xi(r)$ and
$\xi(r_p,\pi)$ respectively, removing the subscripts $R$ and $S$. 
\begin{figure*}
  \begin{center}
    \includegraphics[width=16cm]{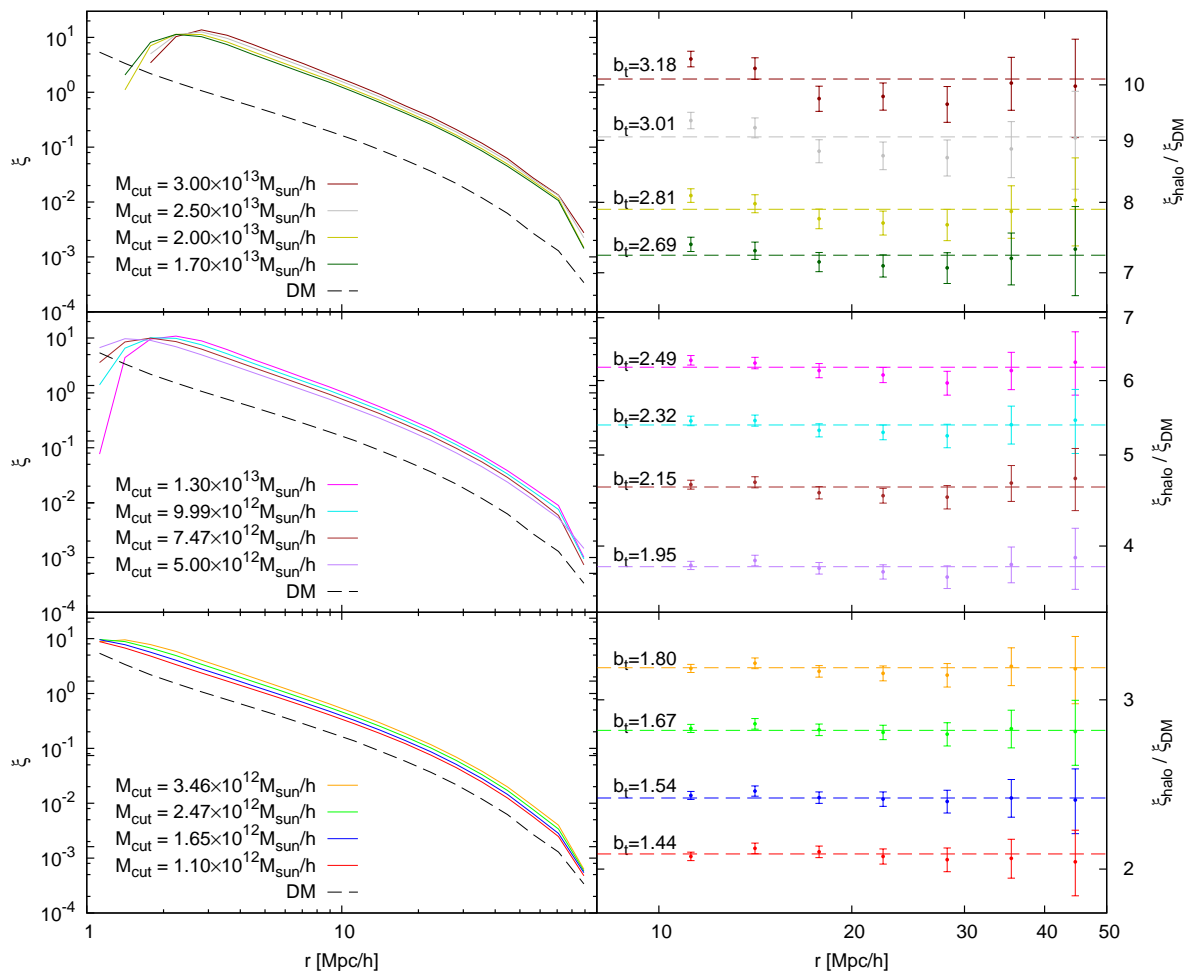}
    \caption{Left: the real-space correlation functions of the
      halo catalogues, compared to that of the dark-matter particles
      in the BASICC simulation. 
    Right: the ratio of $\xi_{halo}(r)$ and $\xi_{DM}(r)$ for
    each catalogue, with the resulting best-fit linear bias  $b_t^2 = \xi_{halo}(r) / 
    \xi_{DM}(r) = \text{const}$, fitted over the range $10<r<50 \, h^{-1} \, \text{Mpc}$.
    Error bars correspond to the standard deviation (of the mean) over
    $27$ sub-cubes.}
    \label{fig xi}
  \end{center} 
\end{figure*} 

\subsection{Fitting the redshift-space correlation function \label{sec fitting procedure}}
We can estimate $\beta$ (and
$\sigma_{12}$, for the linear-exponential model)  through this
modelling, by minimizing the following $\chi^2$ function over a spatial grid:
\begin{equation}
 \label{eq fit}
  \chi^2 = -2 \ln \mathcal{L} = \sum_{i,j} \frac{{(y_{ij}^{(m)} - y_{ij})}^2}{\delta_{ij}^2} \; ,
\end{equation}
where $\mathcal{L}$ is the likelihood and we have defined the quantity 
\begin{equation}
 \label{eq log fit}
 y_{ij} = \log [1 + \xi({r_p}_i,\pi_j)] \; .
\end{equation}
Here the superscript $m$ indicates the model and $\delta^2_{ij}$
  represents the variance of $y_{ij}$. 
The use of $\log (1 + \xi)$ in Eq. (\ref{eq log fit}) has the
advantage of placing more 
weight on large (linear) scales (\citealt{2003MNRAS.346...78H}).
However, unlike \citet{2003MNRAS.346...78H}, we simply use the sample variance of
$y_{ij}$ to estimate $\delta_{ij}$ (as in \citealt {2008Natur.451..541G}). 
We show in Appendix \ref{sec likelihood} that this definition 
provides more stable estimates of $\beta$ also in the low-density
regime. 
The correlation functions are measured using the minimum variance
estimator of \citet{1993ApJ...412...64L}.  We tested different
estimators, such as \citet{1983ApJ...267..465D},
\citet{1982MNRAS.201..867H} and \citet{1993ApJ...417...19H}, finding
that our measurements are virtually insensitive to the estimator
choice, at least for $r \lesssim 50 \ h^{-1} \ \text{Mpc}$.
For the linear-exponential model, we perform a two-parameter fit, 
  including the velocity dispersion, $\sigma_{12}$, as a free
  parameter.  However, being our interest here focused on measurements
  of the growth rate (through $\beta$), $\sigma_{12}$ is treated
  merely as an extra parameter to (potentially) account for deviations
  from linear theory\footnote{See, for instance,
    \citet{2004PhRvD..70h3007S} for a detailed discussion about the 
   physical meaning of $\sigma_{12}$.}. 

Finally, in performing the fit we have neglected an important aspect,
but for good reasons.  In principle, we should consider that the bins of the
correlation function are not independent.  As such, Eq.~(\ref{eq fit})
should be modified as to include also the contribution of non-diagonal terms in the
covariance matrix, i.e. (in matrix form)
\begin{equation}
 -2 \ln \mathcal{L} = {\left(\Ymat^{(m)}-\Ymat\right)}^T \Cmat^{-1}
{\left(\Ymat^{(m)}-\Ymat\right)} \; ,
\end{equation}
where $\Ymat$ and $\Ymat^{(m)}$ are two (column) vectors containing all data and model values respectively
(with dimension $N_b^2$, where $N_b$ is the number of bins in one dimension used to
estimate $\xi(r_p,\pi)$), whereas $\Cmat$ is the covariance matrix, with
dimension $N_b^2\times N_b^2$. 

This is routinely used when fitting 1D correlation functions (e.g.
\citealt{1994MNRAS.266...50F}), but it becomes arduous in the case of
the full $\xi(r_p,\pi)$, for which $N_b \approx 100$ and the
covariance matrix has $\approx 10^8$ elements. What
happens in practice, is that the estimated functions are over-
sampled, so that the effective number of degrees of freedom in the
data is smaller than the number of components in the covariance
matrix, which is then singular.
Still, a test with as many as 100 blockwise boostrap realizations yields
a very unsatisfactory covariance matrix. We tested on a smaller-size
$\xi(r_p, \pi)$ the actual effect of assuming negligible off-diagonal elements
in the covariance matrix, obtaining a difference of a few percent in the
measured value of $\beta$, as also found in \citet{2012arXiv1202.5559D}.
Part of this insensitivity is certainly related to the very 
large volumes of the mock samples, with respect to the scales involved
in the parameter estimations. This
corroborates our forced choice of ignoring covariances in the present
work, also because of the computational time involved in inverting
such large matrices, size multiplied by the huge number of estimates
needed for the present work.

\subsection{Reference distortion parameters and bias values of the simulated samples \label{sec b}}
\begin{figure} 
 \begin{center}
   \includegraphics[width=8cm]{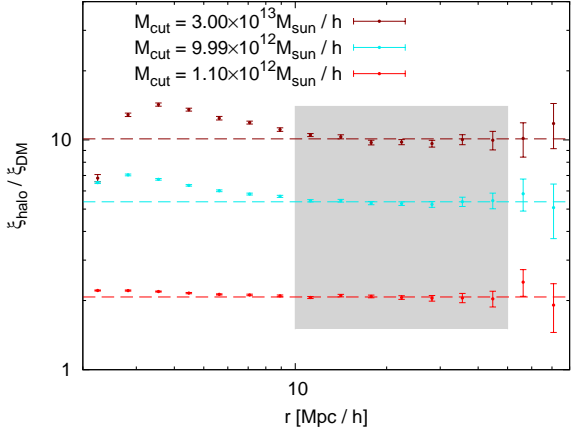}
   \caption{The expected the bias factor, expressed as $b^2=\xi_{halo}(r)
     / \xi_{DM}(r)$, plotted over a wider range of separations than in
     the previous figure.  
   Dashed lines are obtained by fitting a constant bias model over the
   range denoted by the grey area, $10<r<50 \, h^{-1} \, \text{Mpc}$. 
   Error bars give the standard deviation of the mean over the 27 sub-cubes.}
  \label{fig b sq}
 \end{center}
\end{figure}
Before measuring the amplitude of redshift distortions
in the various samples described above, we need to establish the
reference values to which our measurements will be compared, in order to identify systematic effects.  
Specifically, we need to determine with the highest possible accuracy
the intrinsic ``true'' value of $\beta$ for all mass-selected samples in the simulation. 
This can be obtained from the relation \citep{1980lssu.book.....P, 1985PhLB..158..211F, 1990ApJS...74..831L, 1998ApJ...508..483W}
\begin{equation} \label{eq beta true}
 \beta(z) = \frac{\Omega_M^{0.55}(z)}{b(z)} \; ,
\end{equation} 
where, $f(z)=\Omega_M^{0.55}(z)$ is the growth rate of fluctuations at the given
redshift\footnote{In this section we adopt the notation $\Omega_M=\Omega_M(z)$ and $\Omega_{M0}=\Omega_M(z=0)$, not to be confused with the notation $\Omega_M=\Omega_M(z=0)$ adopted elsewhere in this work.}.
For the flat cosmology of the simulation $\Omega_M(z)$ is
\begin{equation}
 \Omega_M(z) = \frac{{(1 + z)}^3 \Omega_{M0}}{{(1 + z)}^3 \Omega_{M0} + (1 - \Omega_{M0})} \; .
\end{equation}
The linear bias can be estimated as
\begin{equation} 
 b^2 = \dfrac{\xi_{halo}(r)}{\xi_{DM}(r)} \; .
 \label{eq xi ratio}
\end{equation}
Here $\xi_{halo}$ and $\xi_{DM}$ have to be evaluated at large separations,
$r \gtrsim 10 \, h^{-1} \, \text{Mpc}$, where the linear approximation
holds.  In the following we shall adopt the notation $b_t$ and
$\beta_t$ for the values thus obtained.  To recover the bias and its
error for each $M_{cut}$ listed in Table \ref{tab halo masses} we
split each cubic catalogue of halos into 27 sub-cubes.  Figure
\ref{fig xi} shows the measured two-point correlation functions and
the corresponding bias values for the various sub-samples. 
These are computed at different separations
$r$, as the average over 27 sub-cubes, with error bars corresponding to
the standard deviation of the mean.
Dashed lines give the corresponding value of $b_t^2$, obtained by
fitting a constant over the range $10<r<50 \, h^{-1} \, \text{Mpc}$.
In most cases, the bias functions show a similar scale dependence, but
the fluctuations are compatible with scale-independence within the
error bars (in particular for halo masses $M_{cut} \le 1.70  \times
10^{13} \ h^{-1} \ M_\odot$). 
For completeness, in Figure \ref{fig b sq} we show that this
remains valid on larger scales ($r\gtrsim 50 \, h^{-1} \, \text{Mpc}$,
whereas on small scales ($r\lesssim 10 \, h^{-1} \, \text{Mpc}$), a
significant scale-dependence is present.  The linear bias assumption
is therefore acceptable for $r\gtrsim 10 \, h^{-1} \, \text{Mpc}$.  

\begin{figure}
 \begin{center}
   \includegraphics[width=8cm]{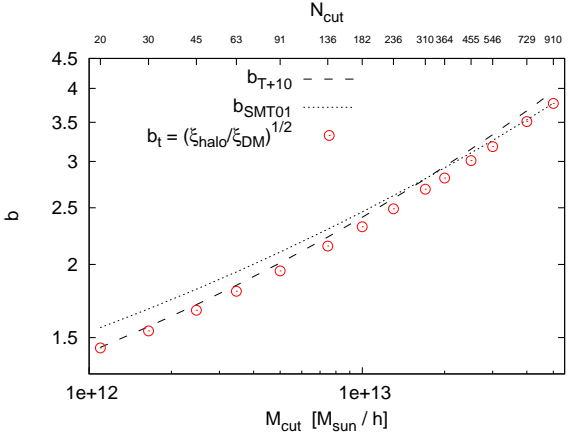}
    \caption{Comparison of the bias values measured from the
      simulated catalogues as a function of their threshold mass,
      $M_{cut}$, with the predictions of the SMT01 and T+10 models.
    The top axis also reports the number of particles per halo,
    $N_{cut}$, corresponding to the catalogue threshold mass.} 
  \label{fig b vs m}
 \end{center}
\end{figure}

In a realistic scenario, $\beta$ is measured from a redshift survey.
Then the growth rate is recovered as $f = b \beta$.
Unfortunately in a real survey it is not possible to estimate $b$
through Eq. (\ref{eq xi ratio}) as we described above (and as it is
done for dark matter simulations) since the real observable is the
two-point correlation function of \emph{galaxies}, whereas $\xi_{DM}$
cannot be directly observed. 
A possible solution is to assume a model for the dependence of the bias on the mass.
Using groups/clusters in this context may be convenient as their total
(DM) mass can be estimated from the X-ray emission temperature or
luminosity. 
We compare our directly measured $b$ with those calculated from two
popular models: \citet*{2001MNRAS.323....1S} and
\citet{2010ApJ...724..878T} (hereafter SMT01 and T+10), in Figure
\ref{fig b vs m}. 
Details on how we compute $b_{SMT01}$ and $b_{T+10}$ are reported
  in the parallel paper by \citet{marulli2012}. 
We see that for small/intermediate masses our measurements are in good
agreement with T+10, whereas for larger masses, $M_{cut} \gtrsim 2
\times 10^{13} \ h^{-1} \ M_\odot$, SMT01 yields a more reliable
prediction of the bias. 

\begin{figure}
  \begin{center}
     \vspace{-0.5cm}	
     \includegraphics[width=7.3cm]{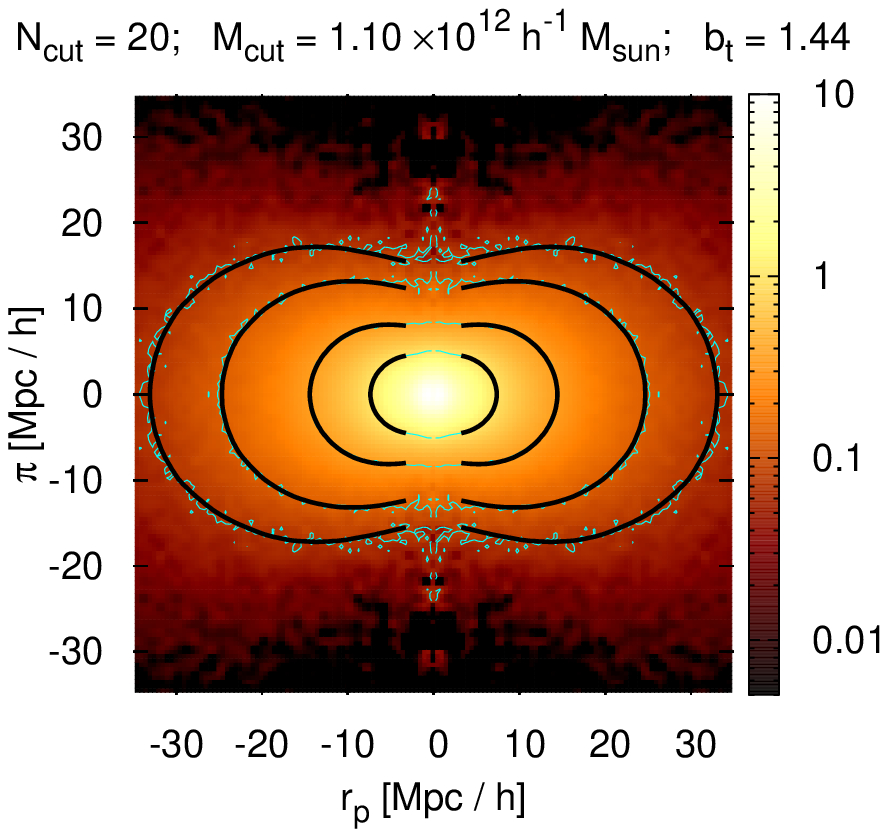} \\
     \vspace{-1cm}
     \includegraphics[width=7.3cm]{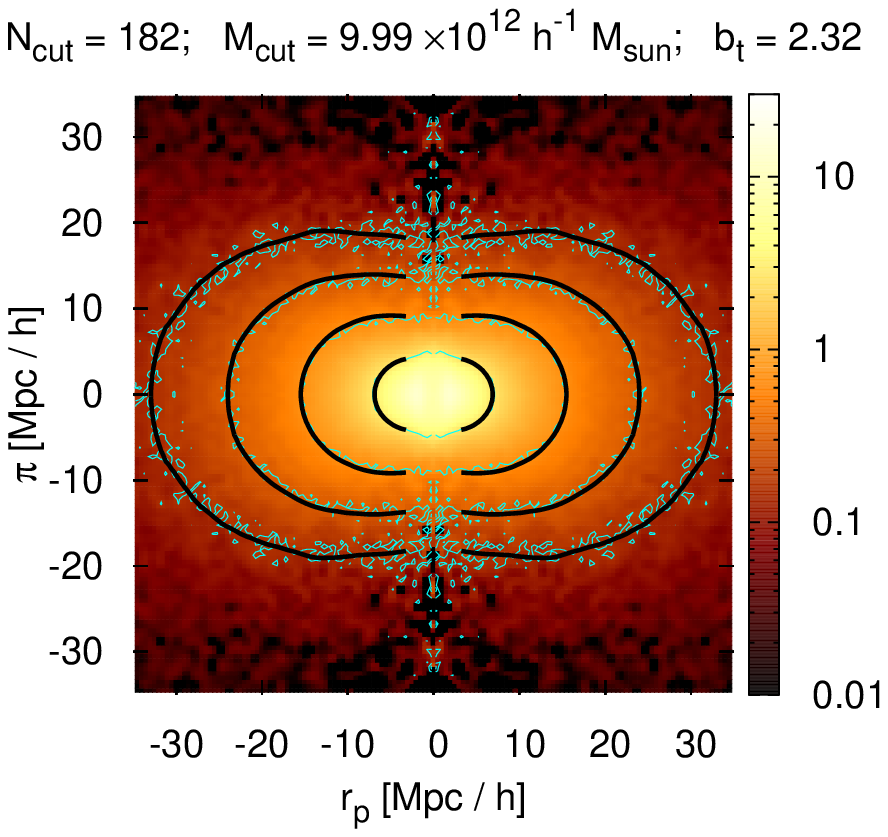} \\
     \vspace{-1cm}
     \includegraphics[width=7.3cm]{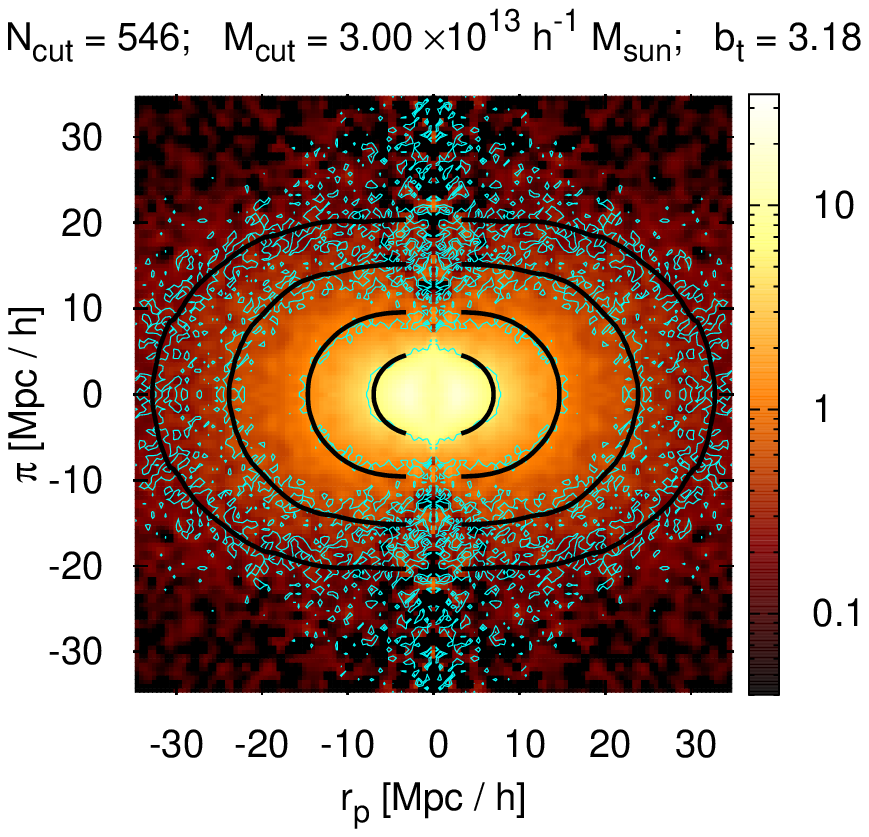}
     \caption{$\xi(r_p,\pi)$ for the catalogues with $M_{cut} = 1.10 \times
       10^{12} \ h^{-1} \ M_\odot$ (upper panel), $M_{cut} = 9.99
       \times 10^{12} \ h^{-1} \ M_\odot$ (central panel) and $M_{cut}
       = 3.00 \times 10^{13} \ h^{-1} \ M_\odot$ (lower panel). 
     Iso-correlation contours of the data are shown in cyan, whereas the
     best fit model corresponds to the black curves. 
    Note that the color scale and contour levels differ in the three panels. The latter
     are arbitrarily set to
     \{0.07, 0.13, 0.35, 1\}, \{0.15, 0.3,  0.7,  2.8\} and \{0.25, 0.5,  1.3,  5\} respectively from top
     to bottom. When the mass grows, the distortion parameter $\beta$ (i.e. the
     compression of the pattern along the line of sight) decreases,
     whereas the correlation and the shot-noise increase.} 
     \label{fig xi bfly}
  \end{center}
\end{figure}

\section{Systematic errors in measurements of the growth rate} \label{sec err}

\subsection{Fitting the linear-exponential model}
As in the previous section, we split each of the 12 mass-selected halo
catalogues of Table~\ref{tab halo masses} into 27 sub-cubes. 
Then we compute the redshift-space correlation function $\xi(r_p,\pi)$
for each of them. Figure \ref{fig xi bfly} gives an example of three
cases of different mass. 
Following the procedure described in Section~\ref{sec fitting
  procedure}, we obtain an estimate of the distortion parameter
$\beta$. 
The 27 values of $\beta$ are then used to estimate the mean value and
standard deviation of $\beta$ as a function of the mass threshold
(i.e. bias).
With the adopted setup (binning and range), the fit becomes unstable for $M_{cut} > 3 \times 10^{13} \ h^{-1} \ M_\odot$, in the sense of yielding highly fluctuating values for $\beta$ and its scatter.
Very probably, this is due to the increasing sparseness of the samples and the reduced amplitude of the distortion (since $\beta \propto 1/b$). 
Figure \ref{fig xi bfly} explicitly shows these two effects: when the
mass grows (top to bottom panels) the shot-noise, which depends on the
number density, increases, whereas the compression along the line of
sight decreases, since it depends on the amplitude of $\beta$. 
For this reason, in this work we consider only catalogues below this mass
threshold, as listed in Table~\ref{tab halo masses}.

\begin{figure*}
  \begin{center}
    \includegraphics[width=16cm]{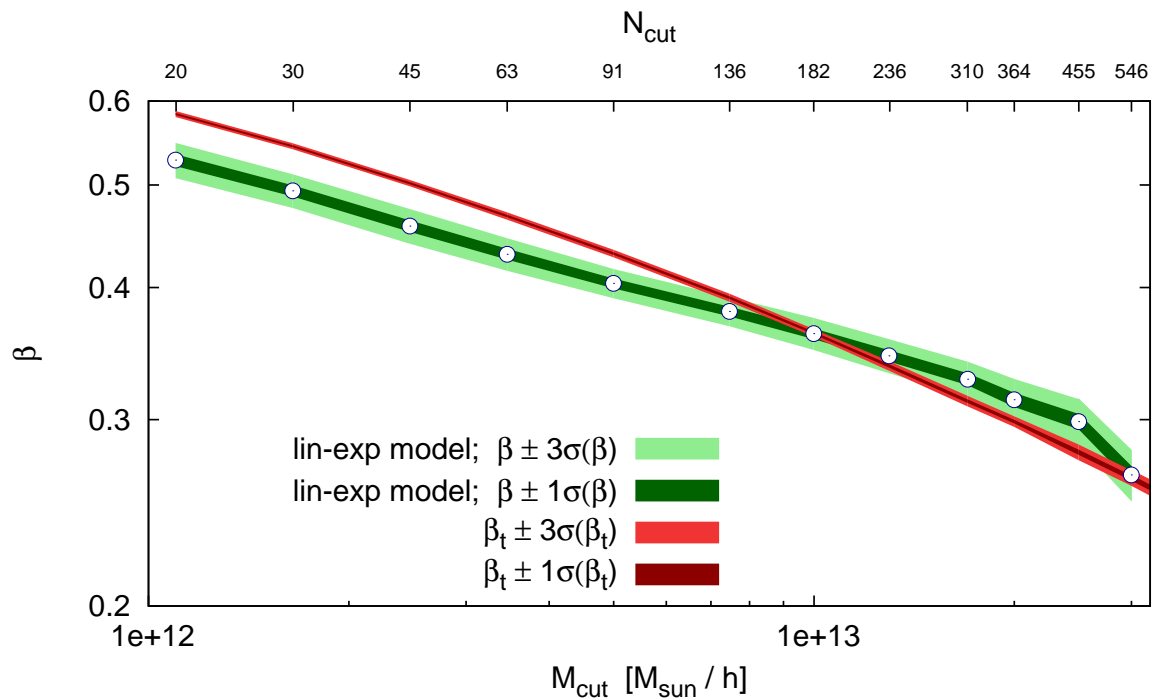}
    \caption{The mean values of $\beta$ averaged over 27 sub-cubes, as
      measured in each mass sample (open
      circles) estimated using the ``standard''  linear-exponential model of
      Eq. (\ref{eq  lin-exp model}). The dark- and light-green bands
      give respectively the  $1 \sigma$ and $3 \sigma$ confidence
      intervals around the mean. The measured values are compared to
      the expected values $\beta_t$, 
      computed using Eqs. (\ref{eq beta  true}-\ref{eq xi
        ratio}). We also give the $1\sigma$ and $3 \sigma$ theoretical
      uncertainty around
      $\beta_t$, due to the uncertainty in the bias estimate (
      brown and red bands, respectively).   
} 
    \label{fig beta lin-exp}
  \end{center}
\end{figure*}

Figure \ref{fig beta lin-exp} summarizes our results.  The plot shows
the mean values of $\beta$ for each mass sample, together with their
confidence intervals (obtained from the scatter of the sub-cubes),
compared to the expected values of the simulation $\beta_t$ (also
plotted with their uncertainties, due to the error on the measured
bias $b_t$, Section~\ref{sec b}).  These have been obtained using the
linear-exponential model, Eq. (\ref{eq lin-exp model}), which
represents the standard approach in previous works, fitting over the
range $3 < r_p < 35 \, h^{-1}\text{Mpc}$, $0 < \pi < 35 \,
h^{-1}\text{Mpc}$ with linear bins of $0.5 \, h^{-1}\text{Mpc}$.  We
also remark that here the model is built using the ``true'' $\xi(r)$
measured directly in real-space, which is not directly observable in
the case of real data.  This is done as to clearly separate the
limitations depending on the linear assumption, from those introduced
by a limited recontruction of the underlying real-space correlation
function.  In Appendix \ref{sec deprojected} we shall therefore discuss
separately the effects of deriving $\xi(r)$ directly from the observations. 

Despite the apparently very good fits (Fig. \ref{fig xi bfly}), we find
a systematic discrepancy between the measured and the true value of
$\beta$.  The systematic error is maximum ($\approx 10\%$) for low-bias
(i.e. low mass) halos and tends to decrease for larger values (note
that here with ``low bias'' we indicate galaxy-sized halos with
$M \approx 10^{12} \ h^{-1} \ M_\odot$).  In particular for $M_{cut}$ between $7 \times
10^{12} $ and $\approx 10^{13} \ h^{-1} \ M_\odot$ the expectation value of the
measurement is very close to the true value $\beta_t$.

It is interesting, and somewhat
surprising, that, although massive halos are intrinsically sparser
(and hence disfavoured from a statistical point of view), the scatter
of $\beta$ (i.e. the width of the green error corridor in Figure
\ref{fig beta lin-exp}) does not increase in absolute terms, showing little dependence
on the halo mass.  Since the value of $\beta$ is decreasing, however,
the relative error does have a dependence on the bias, as we shall
better discuss in \S~\ref{sec form}.

\subsection{Is a pure Kaiser model preferable for cluster-sized halos?}
\begin{figure*}
  \begin{center}
    \includegraphics[width=16cm]{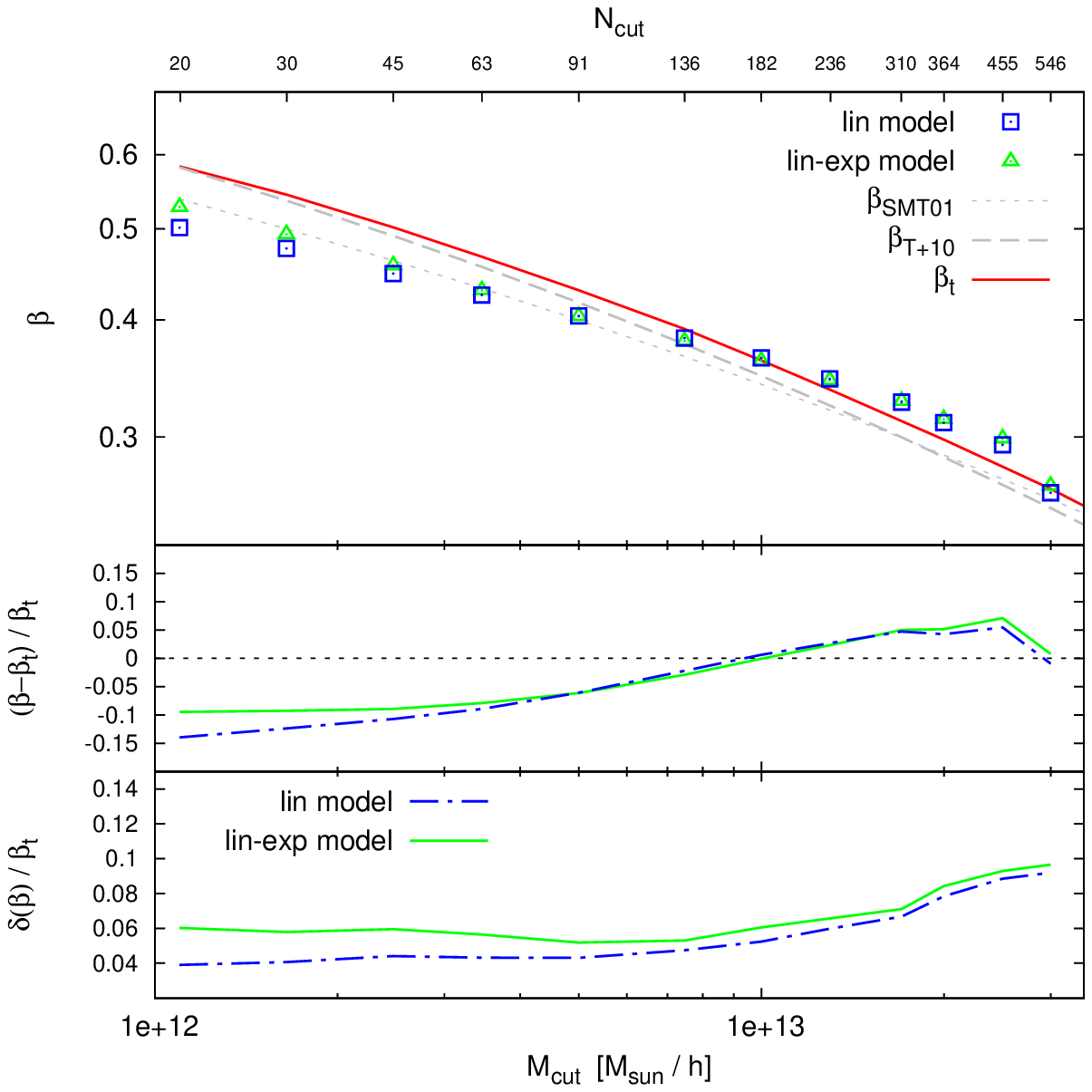}
    \caption{Comparison of the performances of the linear and linear-exponential models. 
    Upper panel: measurements of $\beta$ from the different halo
    catalogues, obtained wth the linear model of Eq. (\ref{eq
      lin model}) (squares) and the linear-exponential model of Eq. (\ref{eq lin-exp model})
    (trianglesl).  Mean values and errors are computed as in
    Fig.~\ref{fig beta lin-exp} from the 27 sub-cubes of each catalogue. 
We also plot the expected values of $\beta$ from the simulation, $\beta_t = f / b_t$ (i.e. $\beta$ 
    \textquotedblleft true\textquotedblright) and from the models of
    Fig.~\ref{fig b vs m}, $\beta_{T+10} = f /
    b_{T+10}$ and $\beta_{SMT01} = f / b_{SMT01}$. 
    Central panel: relative systematic error. 
    Lower panel: relative statistical error.
} 
  \label{fig lin vs exp}
 \end{center}
\end{figure*}

Groups and clusters would seem to be natural candidates to trace
large-scale motions based on a purely linear description, since they
essentially trace very large scales and most non-linear
velocities are confined within their structure. Using clusters as test
particles (i.e. ignoring their internal degrees of freedom) we are
probing mostly linear, coherent motions.  It makes sense therefore to
repeat our measurements using the linear model alone, without
exponential damping correction.   The results are shown in
Figure~\ref{fig lin vs exp}.
The relative error (lower panel) obtained in this case is in general smaller than when the exponential damping is included.
This is a consequence of the fact that the linear model depends only on one free parameter, $\beta$, whereas the linear-exponential model depends on two free parameters, $\beta$ and $\sigma_{12}$.
Both models yield similar systematic
error (central panel), except for the lower mass cutoff range where the
exponential correction clearly has a beneficial effect. 
In the following we briefly summarize how relative and systematic
errors combine. To do this we consider three different mass ranges arbitrarily choosen.
\begin{enumerate}
 \item
   \emph{Small masses} ($M_{cut} \lesssim 5 \times 10^{12} \ h^{-1} M_\odot$) \\
    This range corresponds to halos hosting single $L^*$ galaxies. 
Here the linear exponential model, which gives a smaller systematic
error, is still not able to recover the expected value of $\beta$.  
 However, any consideration about these ``galactic halos'' may not be
 fully realistic since our halo catalogues are lacking in
 sub-structure (see Section \ref{sec Millennium}). 
 \item
   \emph{Intermediate masses}\\
    \hspace*{1.5cm} ($5~\times~10^{12} \lesssim M_{cut} \lesssim 2~\times~10^{13} \ h^{-1}~M_\odot$) \\
    This range corresponds to halos hosting very massive galaxies and
    groups. The systematic error is small compared to that of the other mass
ranges, for both models. 
This means that we are free to use the linear model, which always
gives a smaller statistical error (lower panel), without having to worry too much
about its systematic error, which in any case is not larger than that of the more
complex model. 
In particular, we notice that using the simple linear model in this
mass range, the statistical error on $\beta$ is comparable to that
obtained with a galaxy-mass sample 
using the more phenomenological linear-exponential model. This may be
a reason for preferring the use of this mass 
range for measuring $\beta$. 
 \item
   \emph{Large masses} ($M_{cut} \gtrsim 2 \times 10^{13} \ h^{-1} M_\odot$) \\
    This range corresponds to halos hosting what we may describe as
    large groups or small clusters. 
    The random error increases rapidly with mass (Figure \ref{fig
    lin vs exp}, lower panel), regardless of the model, due to
    the reduction of the distortion signal ($\beta \propto 1/b$)
    and to the decreasing number density.
\end{enumerate}

\subsection{Origin of the systematic errors \label{sec systematics}}
The results of the previous two sections are not fully unexpected.  It
has been evidenced in a number of recent papers that the standard
linear Kaiser description of RSD, Eq. (\ref{eq Kaiser spectrum beta}),
is not sufficiently accurate on the quasi-linear scales ($\approx 5
\div 50 \ \text{h}^{-1} \ \text{Mpc}$) where it is normally applied
(\citealt{2004PhRvD..70h3007S,2006MNRAS.368...85T, 2010arXiv1006.0699T,
  2011MNRAS.410.2081J,
  2011ApJ...726....5O,2011arXiv1105.1194K}).  This involves not only
the linear model, but also what we called the linear-exponential
model.  Since the pioneering work of \citet{1983ApJ...267..465D} the
exponential factor is meant to include the small-scale non-linear
motions, but this is in fact empirical and only partially compensates
for the inaccurate non-linear description. 
The systematic error we quantified with our simulations is thus most
plausibly interpreted as due to the inadequacy of this model on such scales.
Various improved non-linear corrections are proposed in the
quoted papers, although their performance in the case of real galaxies still
requires further refinement (e.g. \citealt{2012arXiv1202.5559D}).
On the other hand, considering larger and larger (i.e. more
linear) scales, one would expect to converge to the Kaiser limit. In
this regime, however, other difficulties emerge, as specifically the
low clustering signal, the need to model the BAO peak
and the wide-angle effects \citep{2011arXiv1102.1014S}.  We have
explored this, although not in a systematic way. We find no indication
for a positive trend in the sense 
of a reduction of the systematic error when increasing the minimum
scale $r_{min}$  included in the fit, at least
for $r_{min}=20 \ h^{-1} \ \text{Mpc}$. Systematic errors remain present, while the statistical
error increases dramatically.  The situation improves only in a
relative sense, because statistical error bars become larger than the systematic
error.  This is seen in more detail in the parallel work by \citet{2012arXiv1202.5559D}.
Finally, it is interesting to remark the indication that systematic errors can be
reduced by using the Kaiser model on objects that are
intrinsically more suitable for a fully linear description. 

\subsection{Role of sub-structure: analysis of the Millennium
  mocks \label{sec Millennium}} 
In the simulated catalogues we use here, sub-structures inside halos, i.e.
sub-halos, are not resolved, due to the use of a single linking length
when running the Friends-of-Friends 
algorithm (Section \ref{sec BASICC}).   As such, the catalogues do not
in fact reproduce correctly the small-scale dynamics observed in real surveys.
Although we expect that our fit (limited to scales $r_p > 3 \ h^{-1} \ \text{Mpc}$)
is not directly sensitive to what happens on the small scales where cluster dynamics dominate,
we have decided to perform here a simple direct check of whether these limitations
might play a role on the results obtained. 
Essentially, we want to understand if the absence of sub-structure
could be responsible for the enhanced systematic error we found for the
low-mass halos. 

To this end, we further analysed $100$ Millennium mock surveys. These
are obtained by combining
the output of the pure dark-matter Millennium run (Springel et al.
2005) with the Munich semi-analytic model of galaxy formation
\citep{delucia07}.  The Millennium Run is a large dark matter
N-body simulation which traces the hierarchical evolution of $2160^3$
particles between $z=127$ and $z=0$ in a cubic volume of
$500^3~h^{-3}~\text{Mpc}^3$, using the same cosmology of the BASICC
simulation $(\Omega_M,~\Omega_\Lambda,~\Omega_b,~h,~n,~\sigma_8)
= (0.25,~0.75,~0.045, ~0.73,~1,~0.9)$. The mass resolution, $8.6
\times 10^8~h^{-1} M_\odot$ allows one to resolve halos containing 
galaxies with a luminosity of $0.1L^*$ with a minimum of 100
particles. Details are given in \citet{springel05}.
The one hundred mocks reproduce the geometry of the VVDS-Wide ``F22''
survey analysed in \citet{2008Natur.451..541G} (except for the fact
that we use complete samples, i.e. with no angular selection
function), covering $2 \times 2 \ \text{deg}^2$ and $0.7 < z < 1.3$.  Clearly,
these samples are significantly smaller than the halo catalogues
built from the BASICC simulations,   
yet they describe galaxies in a more realistic way and allow
us to study what happens on small scales. In addition, while the
BASICC halo catalogues are characterized by a well-defined mass
threshold, the Millennium mocks are meant to reproduce the selection
function of an $I_{AB}<22.5$ magnitude-limited survey like
VVDS-Wide. From each of the 100 light cones, we further consider only
galaxies lying at $0.7<z<1.3$ to have a median redshift close to unity.
The combination of these two sets of simulations should hopefully provide us
with enough information to disentangle real effects from artifacts. 

\begin{figure}
  \begin{center}
    \includegraphics[width=7.3cm]{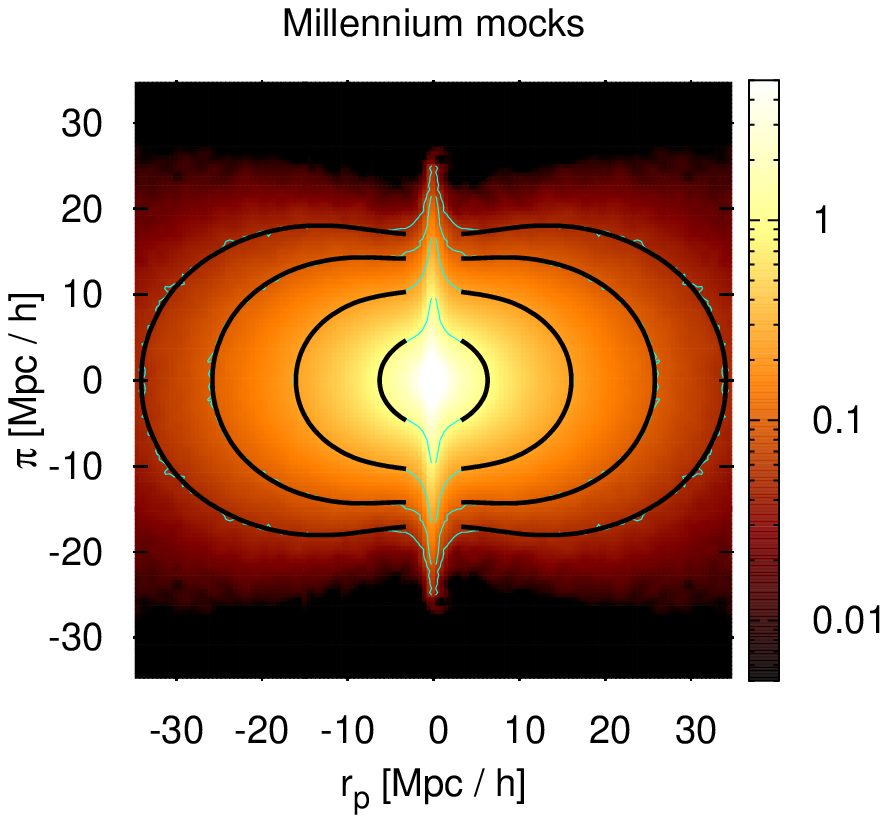}
    \caption{$\xi(r_p,\pi)$ for the Millennium mocks. The coding is the same as in Fig.~\ref{fig xi bfly}, with iso-correlation contours arbitrarily set to \{0.05, 0.1, 0.25, 1\}.} 
  \label{fig MillenniumMocks }
 \end{center}
\end{figure}

Performing the same kind of analysis applied to the BASICC halo catalogues (Figure \ref{fig MillenniumMocks }), we find
a comparable systematic error, corresponding to an under-estimate of $\beta$
by 10\%. We recover $\beta = 0.577 \pm 0.018$, against an 
expected value of $\beta_t = 0.636 \pm 0.006$, suggesting that our
main conclusions are substantially unaffected by the 
limited description of sub-halos in the BASICC samples.  Another potential source of 
systematic errors in the larger simulations could be resolution: the dynamics of the smaller halos 
could be unrealistic simply because they contain too few dark-matter
particles.  Our results from the Millennium mocks and those of
\citet{2011ApJ...726....5O}, which explicitly tested for such effects,
seem however to exclude this possibility.

\section{Forecasting statistical errors in future surveys} \label{sec form}

 A galaxy redshift survey can be essentially characterized by its
 volume $V$ and the number density, $n$, and bias factor, $b$, of
 the galaxy population it includes (besides more specific effects due
 to sample geometry or  selection criteria). 
The precision in determining $\beta$ depends on these parameters.  
Using mock samples from the Millennium run similar to those used here,
\citet{2008Natur.451..541G} calibrated a simple scaling relation for the
relative error on $\beta$, for a sample with $b=1.3$: 
\begin{equation}
  \dfrac{\delta(\beta)}{\beta} \approx \dfrac{50}{n^{0.44} V^{0.5}} \; ,
 \label{eq err Guzzo 2}
\end{equation}
While a general agreement has been found comparing this relation to
Fisher matrix predictions \citep{2009MNRAS.397.1348W}, this formula
was strictly valid for the limited density and volume ranges
originally covered in that work. For example, the
power-law dependence on the density cannot realistically be extended to
arbitrarily high densities, as also pointed out by \citet {2010PhRvD..81d3512S}.  
In this section we present the results of a more systematic
investigation, exploring in more detail the scaling of errors when
varying the survey parameters.  This will include also the dependence
on the bias factor of the galaxy population.  In general, this
approach is expected to provide a description of the error budget
which is superior to a Fisher matrix analysis, as it does not make any
specific assumption on the nature of the errors.  All model fits
presented in the following sections are performed 
using the real-space correlation function $\xi(r)$ recovered from the
``observed'' $\xi(r_p,\pi)$. This is done through the projection/de-projection
procedure described in Appendix~\ref{sec deprojected} (with
$\pi_{max} = 25 \ h^{-1} \text{Mpc}$), which as we show increases the
statistical error by a factor around 2.  The goal here is clearly to
be as close as possible to the analysis of a real data set. 

\subsection{An improved scaling formula \label{sec scaling formula}}

\begin{table*}
  \begin{center}
    \begin{tabular}{cc|ccccccccccccc}
      & & \multicolumn{11}{c}{$n \times 10^5 \, [h^{3} \, \text{Mpc}^{-3}]$}\\

      & & 311 & 204 & 131 & 90.0 & 58.7 & 36.0 & 24.8 & 17.6 & 12.1 & 9.58 & 6.87\\
      \hline

      & $1.10 \times 10^{12}$ & \textopenbullet & \textbullet & \textbullet & \textopenbullet & \textbullet & \textbullet & \textopenbullet & \textbullet & \textbullet & \textbullet & \textbullet & $1.44$\\

      & $1.65 \times 10^{12}$ & & \textbullet & \textbullet & \textbullet & \textbullet & \textbullet & \textbullet & \textbullet & \textbullet & \textbullet & \textbullet & $1.54$\\

      & $2.47 \times 10^{12}$ & & & \textbullet & \textbullet & \textbullet & \textbullet & \textbullet & \textbullet & \textbullet & \textbullet & \textbullet & $1.67$\\

      & $3.46 \times 10^{12}$ & & & & \textbullet & \textbullet & \textbullet & \textopenbullet & \textbullet & \textbullet & \textbullet & \textbullet & $1.80$\\

      & $5.00 \times 10^{12}$ & & & & & \textbullet & \textbullet & \textbullet & \textbullet & \textbullet & \textbullet & \textbullet & $1.95$\\

      $M_{cut} \ [h^{-1} \ M_{\odot}]$ & $7.47 \times 10^{12}$ & & & & & & \textbullet & \textbullet & \textbullet & \textbullet & \textbullet & \textbullet & $2.15$ & $b$\\

      & $9.99 \times 10^{12}$ & & & & & & & \textopenbullet & \textbullet & \textbullet & \textbullet & \textbullet & $2.32$\\

      & $1.30 \times 10^{13}$ & & & & & & & &\textbullet & \textbullet & \textbullet & \textbullet & $2.49$\\
 
      & $1.70 \times 10^{13}$ & & & & & & & & & \textbullet & \textbullet & \textbullet & $2.69$\\

      & $2.00 \times 10^{13}$ & & & & & & & & & & \textbullet & \textbullet & $2.81$\\

      & $2.50 \times 10^{13}$ & & & & & & & & & & & \textbullet & $3.01$\\

     \end{tabular}
  \end{center}
  \caption{Properties of the diluted sub-samples constructed to
    test the dependence of the error of $\beta$ on bias and mean density. 
  Each entry in the table is uniquely defined by a pair
  $(M_{cut},n)$; moving along rows or columns the samples keep a fixed bias
  (mass threshold) or density, respectively.
  Bias values are explicitly reported at the right-hand side of the table.
  The diagonal coincides with the full (i.e. non-diluted) samples. 
  Empty circles indicate catalogues which have been used also to test the dependence on the volume: they have been split into
  $N_{split}^3$ sub-samples for $N_{split} = 3,4,5,6$, whereas all other
  catalogues (filled circles) use $N_{split} = 3$ only for the sake of building
  statistical quantities.} 
 \label{tab dilution}
\end{table*}

In doing this exercise, a specific problem is that, as shown in Table~\ref{tab halo masses},
catalogues with larger mass (i.e. higher bias) are
also less dense.  Our aim is to separate the dependence of the errors
on these two variables. To do so, once a population of a given bias is
defined by choosing a given mass threshold, we construct a series of
diluted samples obtained by randomly removing objects.  The process is
repeated down to a minimum density of $6.87 \times 10^{-5} \ h^3 \
\text{Mpc}^{-3}$, at which shot noise dominates and for the least
massive halos the recovered $\beta$ is consistent with zero.  In this way,
we obtain a series of sub-samples of varying density for fixed bias,
as reported in Table~\ref{tab dilution}.  The full samples are the
same used to build, e.g., Figure \ref{fig beta lin-exp}. 

\begin{figure*}
  \begin{center}
    \includegraphics[width=17cm]{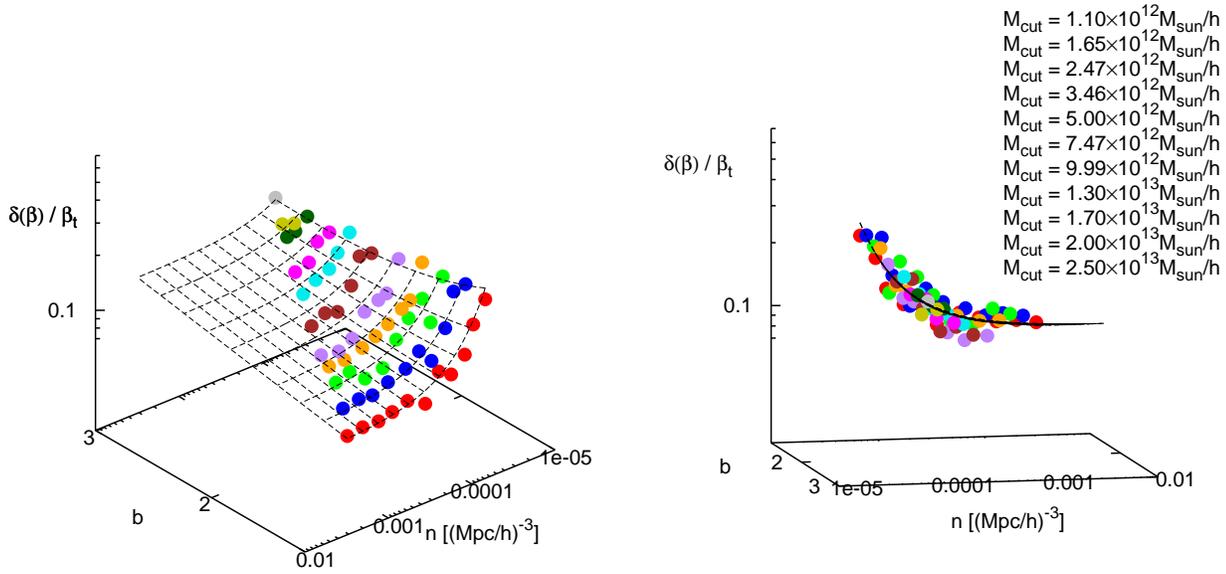}
    \caption{Dependence of the relative error of $\beta$ on the bias and
      number density of the catalogues in Table \ref{tab dilution},
      overplotted on the surface described by the scaling formula of
      Eq. (\ref{eq err mine general}). 
   While the left panel is intended to give an overall view, the right
   panel is expressly oriented to show that the formula 
    is an excellent description of the data.} 
    \label{fig surface}
  \end{center}
\end{figure*}

In Figure \ref{fig surface} we plot the relative errors on $\beta$
measured from each catalogue of Table \ref{tab dilution}, as a function
of the bias factor and the number density. These 3D plots are meant
to provide an overview of the global behavior of the errors; a
more detailed description is provided in Figures \ref{fig
  Fisher n}-\ref{fig Fisher b}, where 2D sections along $n$ and $b$
are reported.  For all the samples considered, the volume is held fixed.

As shown by the figure, the bias dependence is weak and approximately
described by $\delta(\beta)/\beta \propto b^{0.7}$, i.e. the error is
slightly larger for higher-bias objects. This indicates that the gain
of a stronger clustering signal is more than cancelled by
the reduction of the distortion signal, when higher bias objects are
considered.  This is however fully true only for samples which are not too sparse
intrinsically.  We see in fact that at extremely low densities, the
relationship is inverted, with high-bias objects becoming favoured.
 At the same time, there is a clear general flattening of the dependence of the
 error on the mean density $n$. The relation is not a simple
  power-law, but becomes constant at high values of $n$.  In
  comparison, over the density range considered here, the old scaling formula of 
Guzzo et al. would overestimate the error significantly.  This
behaviour is easily interpreted as showing the
transition from a shot-noise dominated regime at low densities to a
cosmic-variance dominated one, in which there is no gain in further
increasing the sampling.  
Such behaviour is clear for low-mass halos (i.e. low bias) but is
much weaker for more massive, intrinsically rare objects.  

\begin{figure*}
  \begin{center}
    \includegraphics[width=17cm]{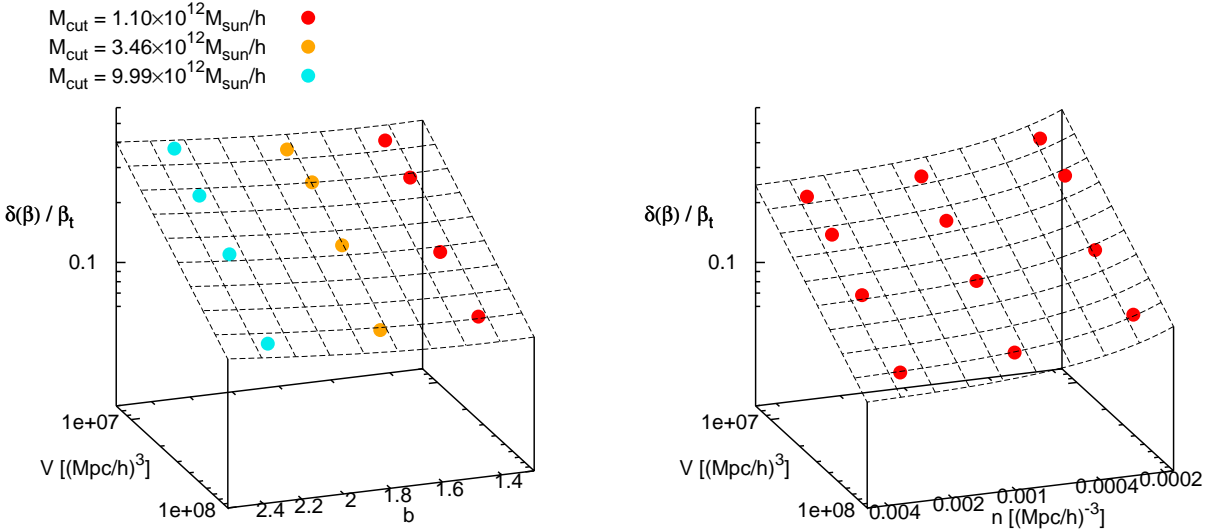}
    \caption{Relative error on $\beta$ as a function of volume, bias
      and number density.  The dependence on volume is explored by
      dividing the sample into 
    $N_{split}^3$ sub-samples, with $N_{split} = 3,4,5,6$. 
    As in all of this section, in modelling the measured
    $\xi(r_p,\pi)$ through Eq.~(\ref{eq lin-exp model})  we use the deprojected
    $\xi(r)$ (with $\pi_{max} = 25 \ h^{-1}\text{Mpc}$), as to
    represent a condition 
    as close as possible to real observations. 
    The superimposed grid is described by the scaling formula of 
    Eq.~(\ref{eq err mine general}). Left panel: $\delta(\beta)
    / \beta_t$ as a function of volume and bias, considering three
    different threshold masses 
    (i.e. biases), but randomly diluting the catalogues as to keep a
    constant number density, $n = 2.48 \times 10^{-4} \, h^{3} \, 
    \text{Mpc}^{-3}$ in all cases (see Table \ref{tab dilution}, empty circles). 
    Right panel: $\delta(\beta) / \beta_t$ as a function of the
    volume, $V$, and the number density, $n$. 
    Here we consider a single threshold mass, $M_{cut} = 1.10 \times
    10^{12} \ h^{-1} \ M_\odot$, corresponding to a constant bias, $b=1.44$.} 
    \label{fig rbt vs vol}
  \end{center}
\end{figure*}

\begin{figure*}
 \begin{center}
  \includegraphics[width=14cm]{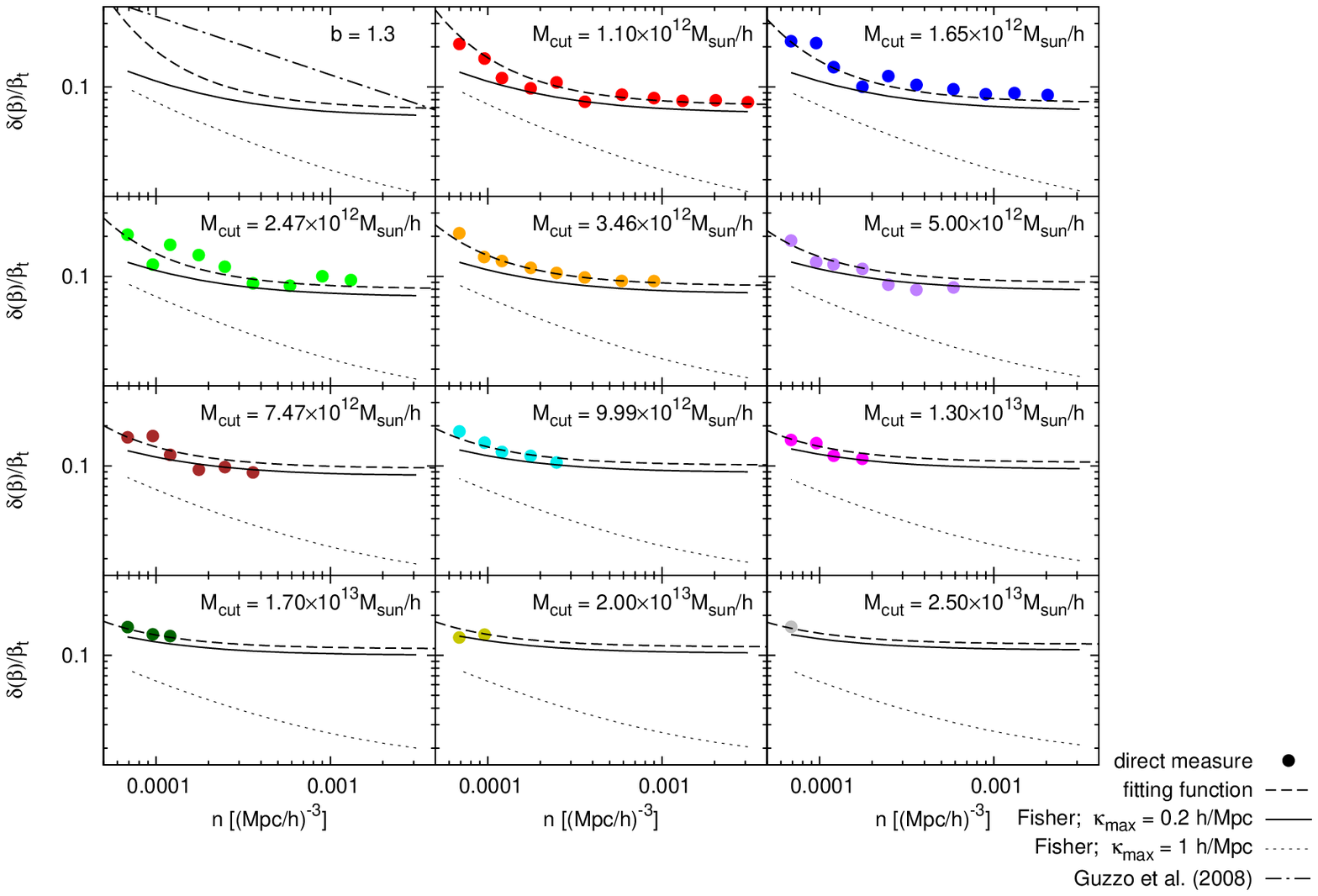}
  \caption{The relative error on $\beta$ as a function
    of the mean number density of the sample, predicted with
    the Fisher matrix approach (solid and dotted lines) and measured
    from the simulated samples (filled circles; colours coded as
    in previous figures). The solid and dotted lines correspond to
    using respectively
    $k_{max}=0.2 \ h \ \text{Mpc}^{-1}$ or $k_{max}=1 \ h \ \text{Mpc}^{-1}$ (with
    Lorentzian damping) in the Fisher forecasts.  
 The dashed lines show in addition the behaviour of the scaling
 formula obtained from the simulation results (Eq.~(\ref{eq err
     mine general})). This is also compared, in the top-left panel, to
   the old simplified fitting formula for $b=1.3$ galaxies of Eq.~(\ref{eq err Guzzo 2}).}
  \label{fig Fisher n}
 \end{center}
\end{figure*}

We can now try to model an improved empirical relation to
reproduce quantitatively these observed dependences.
Let us first consider the general trend, $\delta(\beta)/\beta \propto
b^{0.7}$, which describes well the trend of $\delta(\beta)/\beta$
in the cosmic variance dominated region (i.e. at high density). 
In Figure \ref{fig surface} such a power-law is represented by a plane.
We then need a function capable to warp the plane in the
low density region, where the relative error becomes shot-noise dominated. 
The best choice seems to be an exponential: $\delta(\beta)/\beta
\propto b^{0.7} \exp(n_0/n)$, where, by construction, $n_0$ roughly
corresponds to the threshold density above which cosmic variance
dominates. Finally, we need to add an exponential dependence on the bias so that
at low density the relative error decreases with $b$, such that the
full expression becomes $ \delta(\beta) / \beta \propto  b^{0.7} \exp [n_0/(b^2 n)] $.
The grid shown in Figure \ref{fig surface} represents the result of a
direct fit of this functional form to the data, showing that it is
indeed well suited to describe the overall behaviour.  In the right panel
we have oriented the axes as to highlight the goodness of the fit: 
the {\it rms} of the residual between model and data is 
$\approx 0.015$, which is an order of magnitude smaller than the
smallest measured values of $\delta(\beta)/\beta$. 
This gives our equation the predictive power we were looking for: if
we use it to produce forecasts of the precision of $\beta$ for a given
survey, we shall commit a negligible error\footnote{This estimate is
  obtained by comparing the smallest measured error,
  $\delta(\beta)/\beta \approx 0.07$ (Figure \ref{fig Fisher n}), with
  the {\it rms} of the residuals, $\approx 0.015$.} ($\lesssim 
20\%$) on $\delta(\beta)/\beta$ (at least for values of bias and
volume within the ranges tested here). 
To fully complete the relation, we only need to add the dependence on
the volume, which is in principle the easiest.  To this end, we  
split the whole simulation cube into $N_{split}^3$ sub-cubes,
with $N_{split}=3,4,5,6$. By applying this procedure to 5 samples with
different bias and number density (see Table \ref{tab dilution}) we
make sure that our results do not depend on the particular choice of
bias and density. 
Figure \ref{fig rbt vs vol} shows that $\delta(\beta)/\beta \propto
V^{-0.5}$ independently of $n$ and $b$, confirming the dependence found by
\citet{2008Natur.451..541G}.  
We can thus finally write the full scaling formula for the relative
error of $\beta$ we were seeking for
\begin{equation}
 \delta(\beta) / \beta \approx C b^{0.7} V^{-0.5} \exp \bigg(\dfrac{n_0}{b^2 n}\bigg) \; ,
 \label{eq err mine general}
\end{equation}
where  $n_0=1.7 \times 10^{-4} \ h^3 \ \text{Mpc}^{-3}$ and $C=4.9 \times
10^{2} \ h^{-1.5} \ \text{Mpc}^{1.5}$.  
Clearly, by construction, this scaling formula quantifies random errors, not the systematic ones.

\begin{figure*}
 \begin{center}
  \includegraphics[width=14cm]{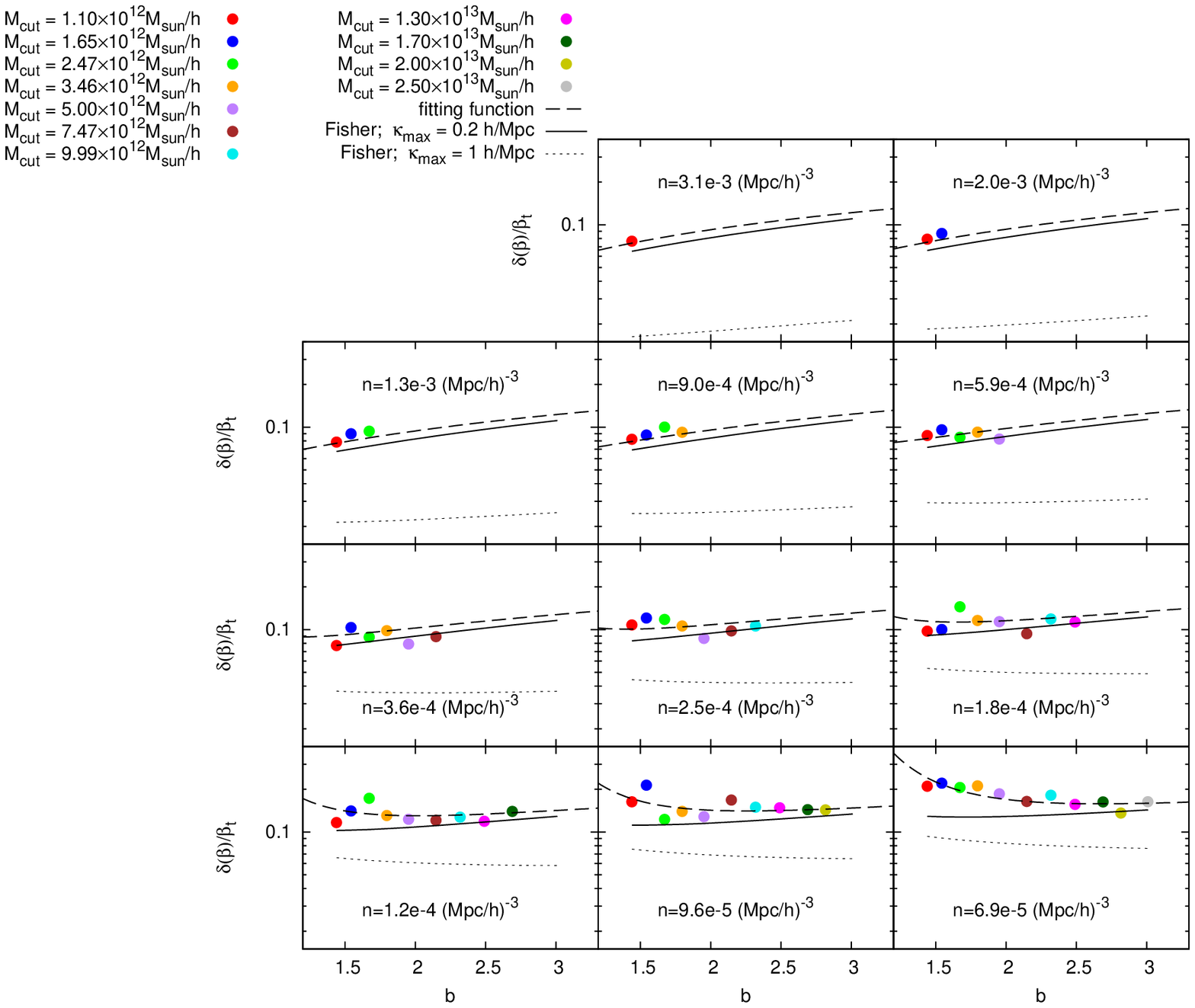}
  \caption{The relative error on $\beta$ as a function
    of the effective bias factor, predicted by
    the Fisher matrix (solid and dotted lines) and measured
    from the simulated samples (filled circles; colours coded as
    in previous figures). The solid and dotted lines correspond to
    using respectively
    $k_{max}=0.2 \ h \ \text{Mpc}^{-1}$ or $k_{max}=1 \ h \ \text{Mpc}^{-1}$ (with
    Lorentzian damping) in the Fisher forecasts.  
 The dashed lines show in addition the behaviour of the scaling
 formula obtained from the simulation results (Eq.~(\ref{eq err
     mine general})).}
  \label{fig Fisher b}
 \end{center}
\end{figure*}

\subsection{Comparison to Fisher matrix predictions}
The Fisher information matrix provides a method for determining the
sensitivity of a particular experiment to a set of parameters and has
been widely used in cosmology. In particular,
\citet{1997PhRvL..79.3806T} introduced an implementation of the Fisher
matrix aimed at forecasting errors on cosmological parameters derived
from the galaxy power spectrum $P(k)$, 
based on its expected observational uncertainty, as
described by \citet[][FKP]{1994ApJ...426...23F}. This was adapted by
\citet{2003ApJ...598..720S} to the measurements of distances using
the baryonic acoustic oscillations in $P(k)$.  Following the renewed
interest in RSD, over the past few years the Fisher matrix technique has also
been applied to predict the errors expected on $\beta$, $f$ and related parameters
\citep[e.g][]{Li2,
  Wn,PW,2009MNRAS.397.1348W, 2010PhRvD..81d3512S, 2010MNRAS.409..737W,
  2011MNRAS.410.1993S, 2011JCAP...10..010B, 2012MNRAS.419..985D}.
The extensive simulations performed here provides us with a natural opportunity to
perform a first simple and direct test of these predictions.  Given
the number of details that enter in the Fisher matrix implementation,
this cannot be considered as exhaustive. Yet, a number of interesting
indications emerge, as we shall see.

We have computed Fisher matrices for all catalogues in Table
\ref{tab dilution}, using a code following
\citet{2009MNRAS.397.1348W}. In particular, our Fisher matrix predicts
errors on $\beta$ and $b$ , given the errors on the linear
redshift space power spectrum modeled as in Eq. (\ref{eq Kaiser
  spectrum beta}) \citep{1987MNRAS.227....1K}.  We first limit the
computations to linear scales, applying the standard cut-off $k <
k_{max} = 0.2 \ h \ {\rm Mpc}^{-1}$. We also explore the possibility 
of including wavenumbers as large as $k=\pi/3\sim 1 \ h \ {\rm
  Mpc}^{-1} $ (that should better match the typical  
scales we fit in the correlation functions from the simulations),
accounting for non-linearity through a conventional
small-scale  Lorentzian damping term. Our fiducial cosmology corresponds to that used
in the simulation, i.e. $\Omega_M = 0.25$, $\Omega_\Lambda = 0.75$,
$H_0 = 0.73$ and $\sigma_8 = 0.9$ today.  We also choose
$\sigma_{12}=200$ km s$^{-1}$ as reference value for the pairwise dispersion.
We do not consider geometric
distortions \citep{1979Natur.281..358A}, whose impact on RSD is 
addressed in the parallel paper by \citet{marulli2012}.
To obtain the Fisher predictions on $\beta$, we marginalize over the
bias, to account for the uncertainty on its precise value, and on the
pairwise velocity in the damping term (when present).

Figure \ref{fig Fisher n} shows the measured relative errors on
$\beta$ as a function of the number density, compared to the
Fisher forecasts for the two choices of $k_{max}$.  We also
plot the scaling relation from Eq. (\ref{eq err mine 
  general}), which best represents the simulation results.
We see that the simulation results are in in fairly good agreement
with the Fisher predictions, when we limit the computation to
very linear scales in the power spectrum (solid line).  The inclusion
of higher wavenumbers
produces unrealistically small errors and with a wrong dependence on
the number density.  Both the solid lines and points reproduce the observed flattening
at high number densities, which corresponds to the transition between a
shot-noise and a cosmic-variance dominated regime, respectively. 

Similarly, Figure \ref{fig Fisher b} looks at the dependence of the
error on the linear bias parameter, comparing the simulation results (points and
scaling formula best-fit) to the Fisher forecasts. 
The behaviour is similar to that observed for the number density:
there is a a fairly good agreement when the Fisher predictions are computed using $
k_{max} = 0.2 \ h \ {\rm Mpc}^{-1}$, except for very low values of the
number density and the bias.  Again, when non-linear scales are
included, the Fisher predictions become too optimistic by a large
factor.

\section{Summary and Discussion} \label{sec concl}

We have performed an extensive investigation of statistical and
systematic errors in measurements of the redshift-distortion parameter
$\beta$ from future surveys. We have considered tracers of the large-scale
distribution of mass with varying levels of bias, corresponding to
objects like galaxies, groups and clusters.  
To this purpose, we have analyzed large catalogues of dark-matter
halos extracted from a snapshot of the BASICC simulation at $z=1$.
Our results clearly evidence the limitations of
the linear description of redshift-space distortions, showing how errors depend on the
typical survey properties (volume and number density) and the
properties of the tracers (bias, i.e. typical mass). Let us recap them
and discuss their main implications.
\begin{itemize}
 \item
  Estimating $\beta$ using the Hamilton/Kaiser
  harmonic expansion of the redshift-space correlation function
  $\xi(r_p,\pi)$ extended to typical scales, leads to a systematic error of up to $10\%$. This is
  much larger than the statistical error of a few percent reachable by
  next-generation surveys. The larger systematic error is found for
  small bias objects, and decreases reaching a minimum  for halos of
  $10^{13} \ h^{-1} \ M_\odot$. This reinforces the trend observed by
  \citet{2011ApJ...726....5O}. 

 \item
 Additional analysis of mock surveys from the Millennium run confirm
 that the observed systematic errors are not the result of potentially
 missing sub-structure in the BASICC halo catalogues. 

\item
  The use of the deprojected correlation function increases
  the statistical error, inducing also some additional systematic
  effects (details are given in Appendix \ref{sec deprojected} and
  also in the companion paper by \citet{marulli2012}). 

\item
  For highly biased objects, which are sparser and whose surveys
  typically cover larger, more linear scales, the simple Kaiser model
  describes fairly well the simulated 
  data, without the need of the empirical damping term with one extra
  parameter accounting for non-linear motions. This results in
  smaller statistical errors. 

\item
  We have derived a comprehensive scaling formula, Eq.~(\ref{eq err mine general}),
  to predict the precision (i.e. relative statistical error) reachable
  on $\beta$ as a function of survey parameters.  
  This expression improves on a previous attempt
  \citep{2008Natur.451..541G}, generalizing the prediction to a population of
  arbitrary bias and properly describing the dependence on the number
  density.  

This formula can be useful to produce quite general
and reliable forecasts for future surveys\footnote{For example, it has
  recently been used, in combination with a Fisher matrix analysis, to
  predict errors on the growth rate expected by the ESA Euclid
  spectroscopic survey [cf. Fig.2.5 of \citet{2011arXiv1110.3193L}]}.
One should in any case consider that there are a few
implementation-specific factors that can modify the absolute values of
the recovered {\it rms} errors.  For example, these would 
depend on the range of scales over which $\xi(r_p,\pi)$ is fitted. The
values obtained here refer to fits performed between $r_{min}=3$ and
$r_{max}=35 \ h^{-1} \ \text{Mpc}$. This has been identified through several
experiments as an optimal range to minimize statistical and systematic
errors for surveys this size \citep{Bianchi2010}.  Theoretically, one may find natural to
push $r_{max}$, or both $r_{min} $ and $r_{max}$ to larger scales, as
to (supposedly) reduce the weight of nonlinear scales.  In practice,
however, in both cases we see that
random errors increase in amplitude (while the systematic error is not
appreciably reduced).  

Similarly, one should also keep in mind that the formula is
strictly valid for $z=1$, i.e. the redshift
where it has been calibrated.  There is no obvious reason to expect the
scaling laws among the different quantities (density, volume, bias) to depend significantly on the
redshift.  This is confirmed by a few preliminary measurements we
performed on halo catalogues from the $z=0.25$ snapshot of the BASICC.
Conversely, the magnitude of the errors may change, as shown, e.g., in \citet{2012arXiv1202.5559D}. 
We expect these effects to be described by a simple renormalization of
the constant $C$.  

Finally, one may also consider that the standard deviations measured using the 27 sub-cubes
could be underestimated, if these are not fully independent.  We
minimize this by maximizing the size of each sub-cube, 
while having enough of them as to build a meaningful statistics. The
side of each of the 27 sub-cubes used is in fact close to $500 \ h^{-1} \ \text{Mpc}$,
benefiting of the large size of the BASICC simulation.

 \item
  We have compared the error estimations from our simulations with idealized
  predictions based on the Fisher matrix approach, customarily implemented in
  Fourier space. We find a good agreement, but only when the Fisher
  computation is limited to significantly large scales, i.e.  $k < k_{max} = 0.2 \ h \ {\rm
    Mpc}^{-1}$.  When more non-linear scales are included (as an
  attempt to roughly match those actually involved in the fitting of $\xi(r_p,
  \pi)$ in configuration space), then the predicted errors become
  unrealistically small.  This indicates that the usual convention of
 adopting $k_{max} \sim 0.2 \ h \ {\rm Mpc}^{-1}$ for these kind of
 studies is well posed. On the other hand, 
 it seems paradoxical that in this way with the two methods we are 
 looking at different ranges of scales.  The critical point clearly lies
 in the idealized nature of the Fisher matrix technique. When moving
 up with $k_{max}$ and thus adding more and more nonlinear scales, the
 Fisher technique simply accumulates signal and dramatically improves
 the predicted error, clearly unaware of the additional ``noise'' introduced by the
 breakdown of linearity.  On the other hand, if in the direct fit
 of  $\xi(r_p, \pi)$ (or $P(k, \mu)$) one conversely considers a corresponding
 very linear range $r>2\pi/k_{max} \sim 30 \ h^{-1} \ \text{Mpc}$, a
 poor fit is obtained, with much larger statistical errors than shown, e.g.,
 in Fig.~\ref{fig beta lin-exp}.  There is no doubt that smaller, mildly nonlinear scales at
 intermediate separations have necessarily to be
 included in the modelling if one aims at reaching percent statistical errors on measurements
 of $\beta$ (or $f$).  If one does this in the Fisher matrix, then the
 predicted errors are too small.
The need to push our estimates to scales which are not
 fully linear will remain true even with surveys of the next
 generation, including tens of millions of galaxies over Gpc volumes,
because that is where the clustering and distortion signals are (and
will still be) the
strongest.  Of course, our parallel results on the amount of systematic
 errors that plague estimates based on the standard dispersion model
 also reinforce the evidence that better modelling of nonlinear
 effects is needed on these scales. The strong
 effort being spent in this direction gives some confidence that significant
 technical progress will happen in the coming years \citep[see
 e.g.][and references therein]{2011arXiv1105.1194K, 2012arXiv1202.5559D}.   

In any case, this limited exploration suggests once more that
forecasts based on the Fisher matrix approach, while giving useful
guidelines evidence the error dependences, have to be treated with 
significant caution and possibly verified with more direct methods.
Similar tension between Fisher and Monte Carlo forecasts has been
recently noticed by \citet{hawken2012}. 

\item
  Finally, in Appendix \ref{sec likelihood} we have also clarified which is the
  most unbiased form to be adopted for the likelihood when fitting
  models to the observed redshift-space correlation function,
  proposing a slightly different form with respect to previous works. 
\end{itemize}

With redshift-space distortions having emerged as probe of primary
interest in current and future dark-energy-oriented galaxy surveys,
the results presented here further stress the need for improved
descriptions of non-linear effects in clustering and dynamical
analyses.  On the other hand, they also indicate the importance of
building surveys for which multiple tracers of RSD (with different bias
values) can be identified and used in combination to help
understanding and minimizing systematic errors.

\section*{Acknowledgments}

We warmly thank M. Bersanelli for discussions and constant support and
C. Baugh for his invaluable contribution to the BASICC
Simulations project.  DB acknowledges support by the Universit\`a
degli Studi di Milano through a PhD fellowship.  EM is supported by
the Spanish MICINNs Juan de la Cierva programme (JCI-2010-08112), by
CICYT through the project FPA-2009 09017 and by the Community of
Madrid through the project HEPHACOS (S2009/ESP-1473) under grant
P-ESP-00346.  Financial support of PRIN-INAF 2007, PRIN-MIUR 2008 and
ASI Contracts I/023/05/0, I/088/06/0, I/016/07/0, I/064/08/0 and
I/039/10/0 is gratefully acknowledged.
LG is partly supported by ERC Advanced Grant \#291521 `DARKLIGHT'.

\appendix

\section{Definition of the likelihood function to estimate $\beta$}
\label{sec likelihood}

To estimate $\beta$, in Section \ref{sec fitting procedure} we defined a
likelihood function comparing the measured correlation 
function $\xi(r_p,\pi)$ and the corresponding parameterized models. Our likelihood is simply
given by the standard $\chi^2$ expression 
\begin{equation}
 \label{eq fit app}
-2 \ln \mathcal{L} =\sum_{i,j} \frac{{(y_{ij}^{(m)} -
        y_{ij})}^2}{\delta_{ij}^2} \; ,
\end{equation}
where however the stochastic variable considered is not just the value of $\xi(r_p,
\pi)$ at each separation $(r_p,\pi)=(r_i, r_j)$, but the expression
\begin{equation}
 y_{ij} = \log [1 + \xi(r_i,r_j)] \; ,
\end{equation}
which has the desirable property of placing more weight on large,
more linear scales. This was first proposed by \citet{2003MNRAS.346...78H},
who correspondingly adopt the following expression for the expectation value
of the variance
\begin{equation}
 \label{eq scatter Hawkins}
 \delta_{ij}^2 = {\{\log [1 + \xi_{ij} + \delta(\xi_{ij})] - \log [1 + \xi_{ij} - \delta(\xi_{ij})]\}}^2 \; .
\end{equation}
This simply maps onto the new variables $y_{ij}$, the interval
including 68\% of the distribution in the original variables $\xi_{ij}$, i.e. twice the standard deviation
if this were Gaussian distributed.
Strictly speaking, here an extra factor 1/2 would be formally required if one aims at defining the equivalent of a standard deviation, but this is in the end uneffective in the minimization and thus in finding the best-fitting parameters.
\begin{figure}
 \begin{center}
  \includegraphics[width=8cm]{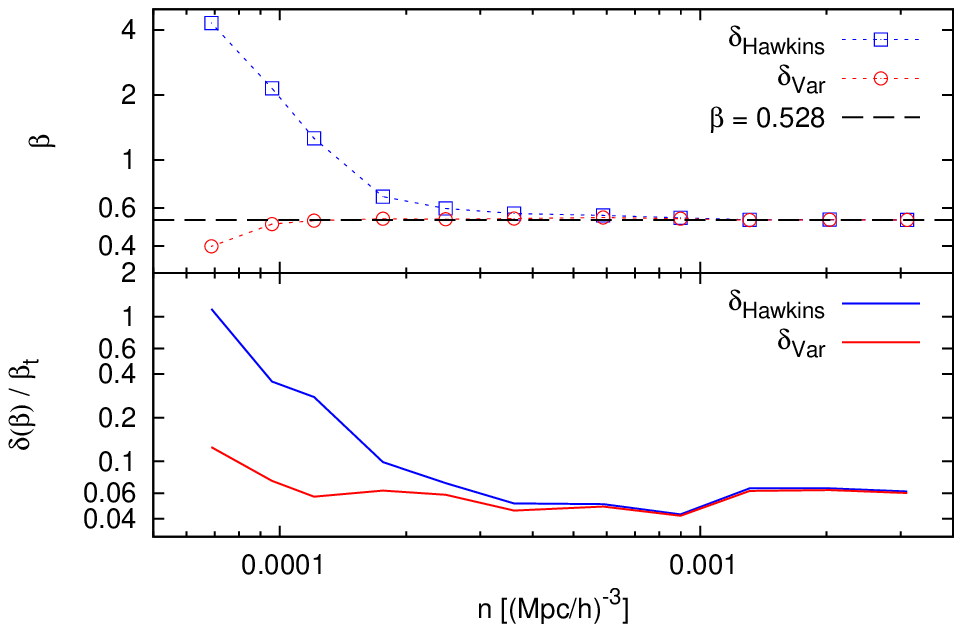}
  \caption{Mean value (top) and relative scatter (bottom)
    of $\beta$, as recovered from catalogues with varying density (but same volume
    and bias), using the two different definitions of the
    variance of each data point of Eqs.~\ref{eq scatter Hawkins} (open
    blue squares) and ~\ref{eq delta y} (open red circles).  The dashed
    line shows as reference the asymptotic common value of $\beta$ that both
    methods identically recover at high densities.
Note how using eq. ~\ref{eq delta y} yields an unbiased estimate down
to significantly smaller densities, whereas the estimator based on
Eq.~(\ref{eq scatter Hawkins}) becomes rapidly more and more biased
below $n \approx 5 \times 10^{-4} \ h^3 \ \text{Mpc}^{-3}$. The intrinsic
scatter of the measurements, as usual obtained from the 
27 sub-cubes of this specific catalogue, also follows a similar trend.
} 
\label{fig bt vs n 20}
 \end{center}
\end{figure}

However, the weighting factors $1/\delta_{ij}$ in the likelihood
  definition depend explicitly on $\xi_{ij}$, which 
may result in an improper weighting of the data when the correlation
signal fluctuates near zero. 
We have directly verified that when the estimate is noisy, it is
preferable to use a smooth weighting scheme rather than one that is
sensitive to local random oscillations of $\xi$, which is more likely
to yield biased estimates.  This supported our choice of adopting the
usual sample-variance expression  
\begin{equation}
 \label{eq delta y}
\delta_{ij}^2 = \frac{1}{N}\sum_{k}  \left(y_{ij}^{(k)} - \left< y_{ij}\right>\right)^2 \; ,
\end{equation}
estimated over $N$ realizations of the survey.
This can be done using mock realizations \citep{2008Natur.451..541G},
or, alternatively, through appropriate jack-knife or booststrap
resamplings of the data. 
Specifically, we find a significant advantage of the weighting
scheme based on sample variance when dealing with low-density samples. 
This is shown in Figure \ref{fig bt vs n 20}, where $\beta$ is estimated on 
the catalogue with $M_{cut} = 1.10 \times10^{12} \ h^{-1} \ M_\odot$ using the
two likelihoods and gradually diluting the sample (note that all computations in
this section use the linear-exponential model, with $\xi(r)$ directly
measured in real-space).  

In order to understand the reasons behind this behaviour, we have
studied independently the various terms composing the likelihood. We
use one single sub-cube (i.e. 1/27 of the total volume),
from the catalogue with $M_{cut} = 1.10 \times 10^{12} \ h^{-1} \ M_\odot$, and 
consider two extreme values of the mean density. 
First, we consider the case of the highest
density achievable by this halo catalogue, $n = 3.11 \times 10^{-3} \, h^3
\, \text{Mpc}^{-3}$.  In the upper panel of Figure \ref{fig enzo perc100
  pi9.75 refHawkins} we plot a section of $\xi(r_p,\pi)$ at constant
$\pi = 9.75 \, h^{-1} \, \text{Mpc}$, together with
the model $\xi_m(r_p,\pi)$ corresponding 
to the best-fit $\beta$ and $\sigma_{12}$ parameters.  In this
density regime the values of the recovered best-fit parameters are essentially independent of the form chosen for
$\delta_{ij}^2$ (as shown by the coincident values of $\beta$ on the right side
of Figure \ref{fig bt vs n 20}).   The match of the model to the data
is very good.  In the central panel, we plot instead, for each bin $i$ along $r_p$,
the absolute value of the difference between model and observation, $\left(|y
  - y_m|\right)_i$, together with the corresponding
standard deviations in the two cases, which are virtually indistinguishable from each other. Finally, the lower panel
shows the full values of the terms contributing to the $\chi^2$ sum, again
showing the equivalence of the two choices in this density regime. 

\begin{figure}
 \begin{center}
  \includegraphics[width=8cm]{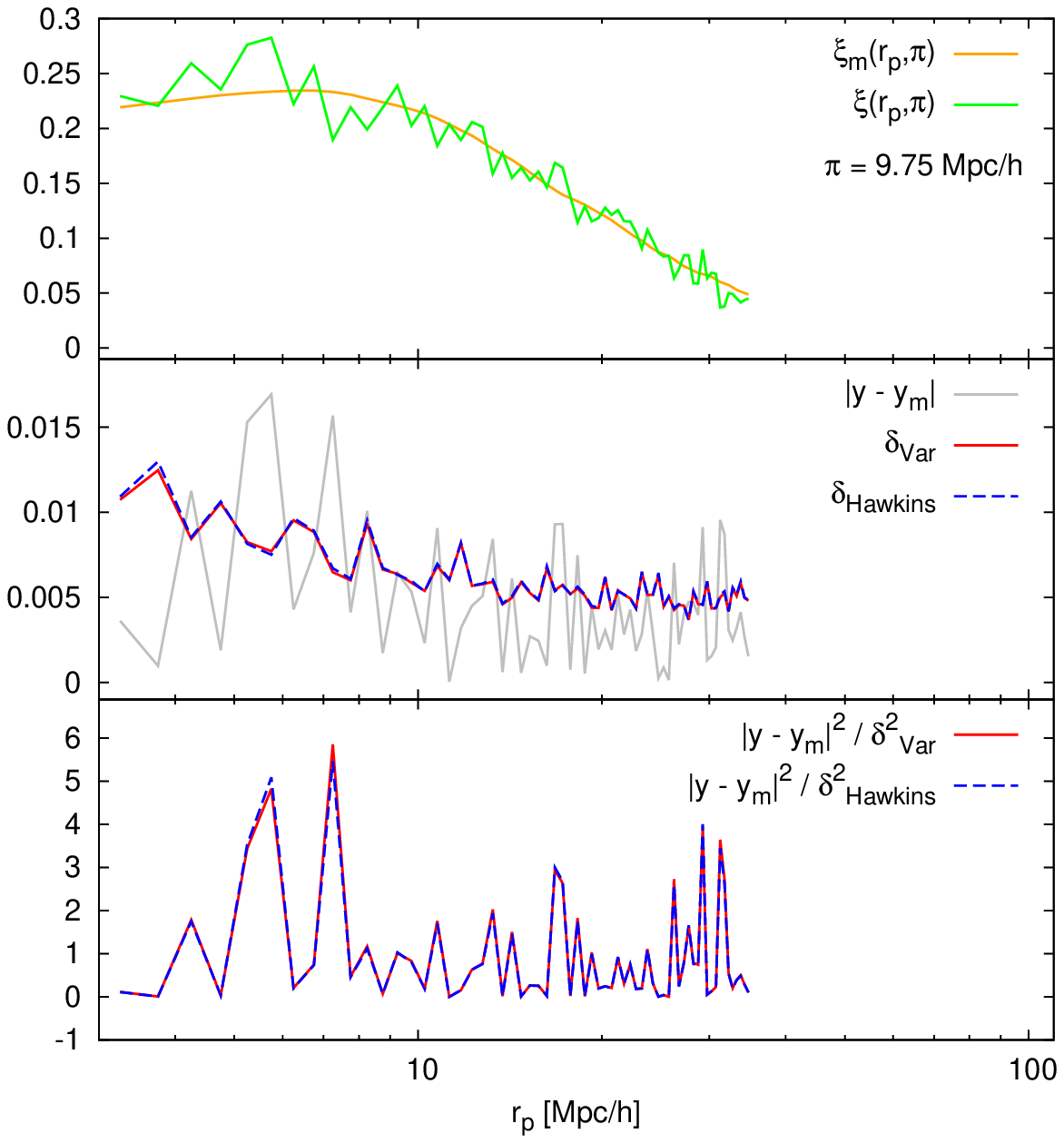}
  \caption{
  Comparison of the performances of the two likelihood forms
    discussed in the text in the high-density regime, using the fully
    sampled population of halos from a single
    sub-cube (1/27 of the volume)  with $M_{cut} = 1.10 \times 10^{12} \ h^{-1} \ M_\odot$.  Top panel: 
    cut-through $\xi(r_p,\pi)$ at fixed  $\pi = 9.75 \, 
  h^{-1} \, \text{Mpc}$ (broken line), and corresponding best
  fit model $\xi_m(r_p,\pi)$ using the Hawkins et al. form for the
  scatter of each data point (continuous line).
  Central panel: residual values $|y_{ij}-y^{(m)}_{ij}|$ between the
  data and model values (light grey line) and values for the scatter
  of each point, according to the two definitions of Eqs. ~\ref{eq delta y}  (solid red line) and ~\ref{eq
    scatter Hawkins} (dashed
  blue line).
  Bottom panel: corresponding terms in the $\chi^2$ sum
  (see Eq. (\ref{eq fit app})).  The two definitions for the scatter,
  as expected, produce virtually identical values for the likelihood.
}
  \label{fig enzo perc100 pi9.75 refHawkins}
\end{center} \end{figure} 

\begin{figure} 
 \begin{center}
  \includegraphics[width=8cm]{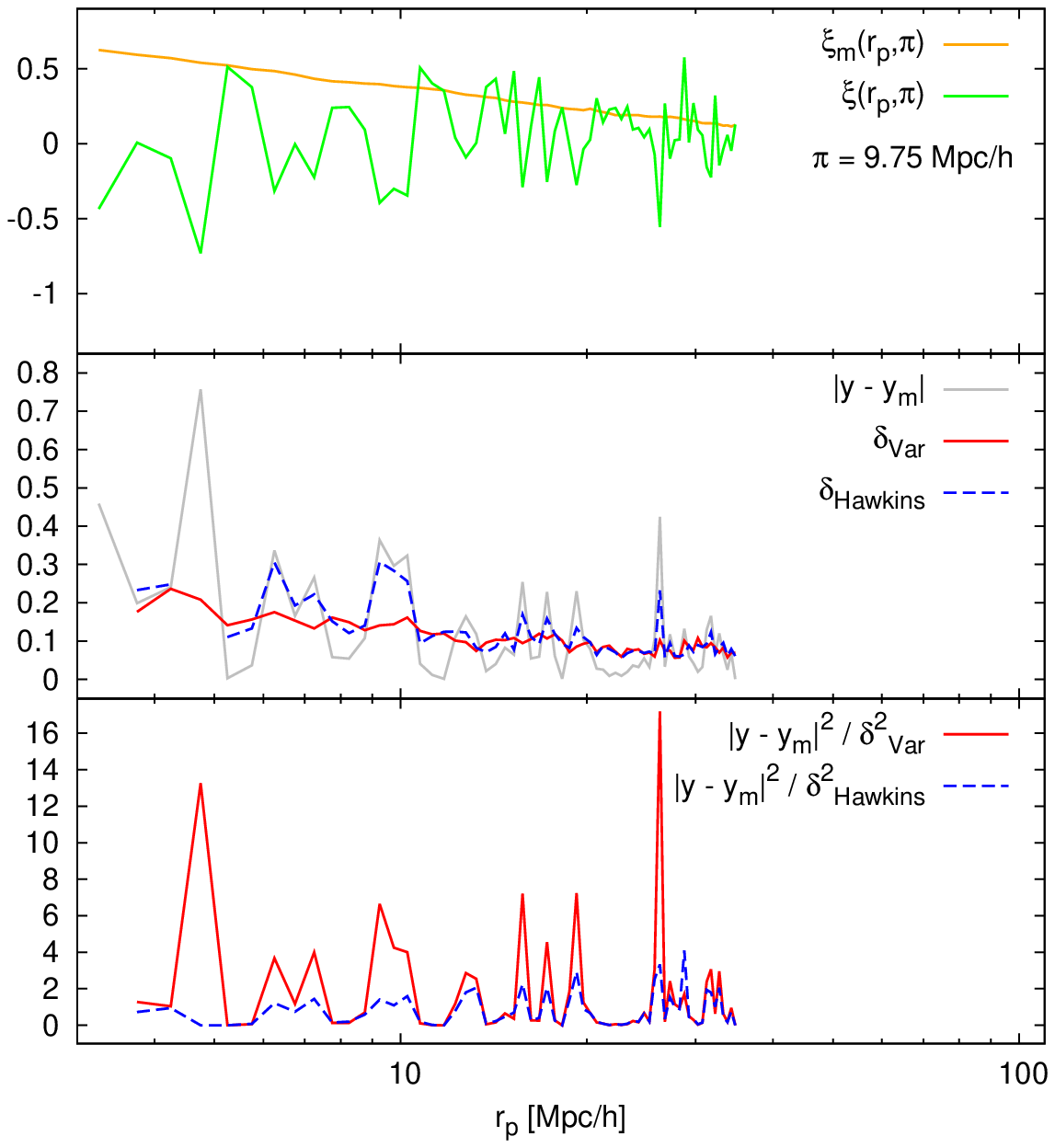}
  \caption{
  Same as Figure \ref{fig enzo perc100 pi9.75 refHawkins}, but now in
  the low-density regime ($n = 9.58 \times 10^{-5} \, h^3 \, \text{Mpc}^{-3}$).
 Again, the model curve in the top panel corresponds to the best-fit
 parameters obtained using the
 Hawkins et al. form of the scatter of each measurements. The fit is
 very unsatisfactory.  The bottom panel shows how the
 likelihood expression based instead on the standard deviation 
 of $y$ as from Eq.~(\ref{eq delta y}) 
 rejects these parameter values, giving high
 $\chi^2$ values (red solid curve).  Note the
 different scale on the ordinate, with respect to previous figure.
}
  \label{fig enzo perc3.08 pi9.75 refHawkins}
 \end{center}
\end{figure}

However, when we sparsely sample the catalogue, as to reach a mean
density of $n = 9.58 \times 10^{-5}$ $h^3 \,
\text{Mpc}^{-3}$ (leaving all other parameters unchanged), a very 
different behaviour emerges (Figure \ref{fig enzo perc3.08 pi9.75 refHawkins})\footnote{In Figure \ref{fig bt vs n 20} (upper panel, second blue square from the left) we show the same behaviour when averaged over 27 sub-samples.}. 
Using the Hawkins et al. definition for the variance yields a best-fit model that overestimates
the data on almost all scales (top panel), corresponding to unphysical
values of $\beta =2.33$ and $\sigma_{12} = 2112 \, \text{km} \, s^{-1}$.
The central panel now shows how in this regime the two definitions of the scatter,
(which weigh the data-model difference), behave
in a significantly different way, with the Hawkins et al. definition being much
less stable than the one used here, and in general anti-correlated with the values
of $\xi(r_p,\pi)$  in the upper panel. 
In the lower panel, the dashed line shows how this anti-correlation
smooths down the $(|y - y_m|)_i$ peaks resulting in erroneously low
values for the $\chi^2$ that drive the fit to a wrong region of the
parameter space. 
In the same panel, the solid line shows how the likelihood computed
with our definition for these same parameters 
gives high $\chi^2$ values, thus correctly rejecting the
model\footnote{For $r_p = 4.75 \ h^{-1} \ \text{Mpc}$ (and $\pi = 9.75
  \, h^{-1} \ \text{Mpc}$) we find $1 + \xi - \delta(\xi) < 0$. 
Consequently, $\delta_{Hawkins}$ is not well defined (Figure \ref{fig
  enzo perc3.08 pi9.75 refHawkins}, central panel) resulting in a zero
weight for the corresponding $\chi^2$ summand (lower panel).}.

\section{Additional systematic effect when using the deprojected correlation function}
\label{sec deprojected}
In a real survey, the direct measurement of $\xi(r)$ is not possible.
A way around this obstacle is to project $\xi(r_p,\pi)$ along the line
of sight, i.e. along the direction affected by redshift distortions. 
We hence define the projected correlation function as
\begin{equation}
\label{eq projected}
w_p(r_p) = 2 \int_0^\infty \xi(r_p,\pi) d\pi = 2 \int_{r_p}^\infty \frac{r'\xi(r')dr'}{\sqrt{r'^2-r_p^2}} \; .
\end{equation}
Inverting the integral we recover $\xi(r)$.
More precisely, following \citet{1992MNRAS.258..134S}, we have
\begin{equation} \label{eq deprojected}
\xi(r) = \dfrac{1}{\pi} \int_r^\infty \dfrac{dw_p (r_p) / d r_p}{\sqrt{r_p^2 - r^2}} d r_p \; ,
\end{equation}
where $\pi$ is the usual mathematical constant, not to be confused
with the line-of-sight separation $\pi$ in Eq. (\ref{eq projected}).

\begin{figure}
  \begin{center}
    \includegraphics[width=8cm]{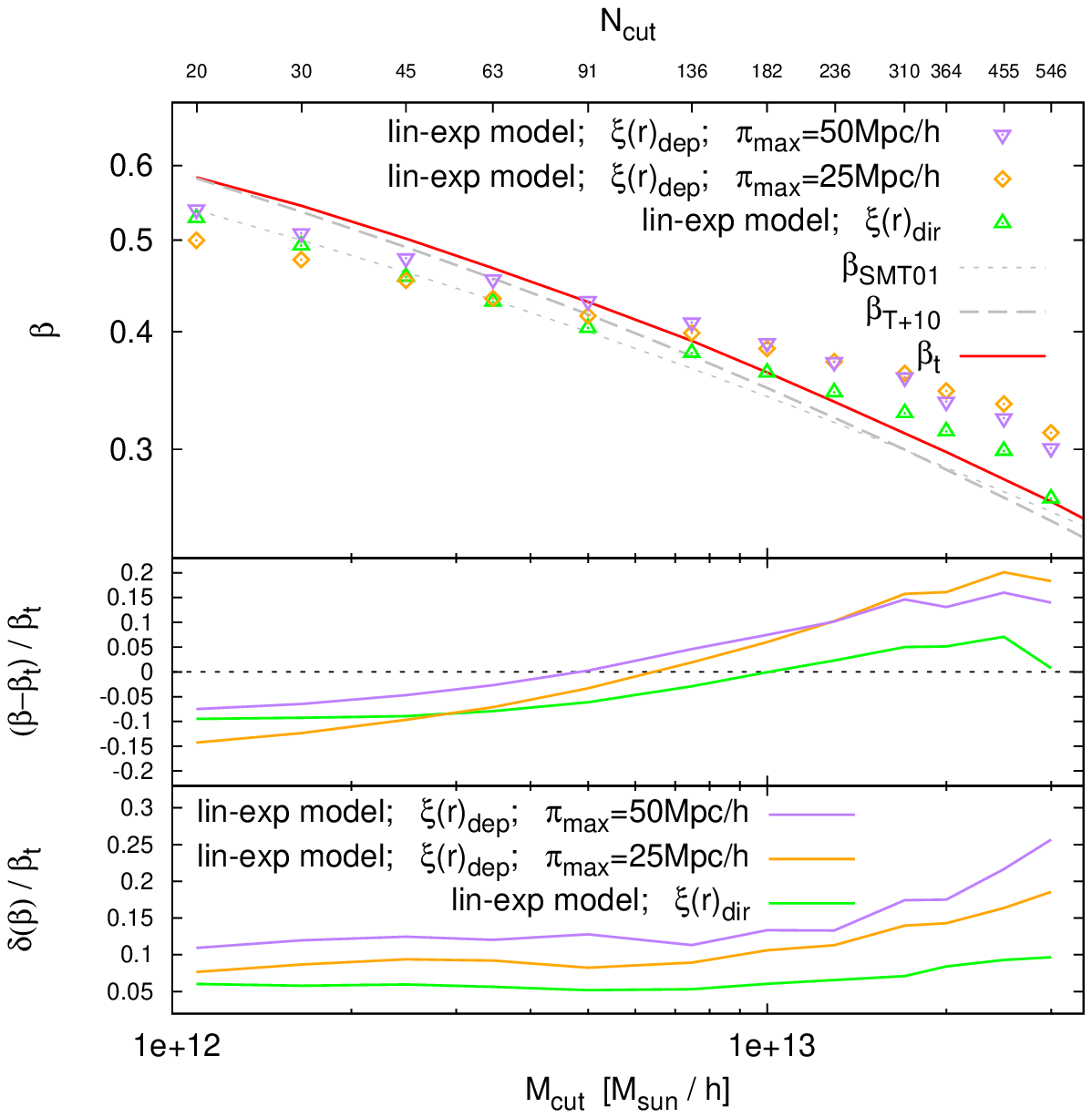}
    \caption{The effect of using the de-projected real-space
      correlation function in the RSD model.
    Upper panel: values of $\beta$ obtained when the real-space
    correlation function $\xi(r)$ is directly measured from the
    simulation (triangles) or deprojected as in real surveys (rhombs
    and inverted triangles). The latter correspond to two different
    integration limits $\pi_{max}$ in the projection.  The two lower
    panels give ths systematic and statistical error as in Figure~\ref{fig lin vs exp}.}
    \label{fig beta dep}
  \end{center}
\end{figure}

A more extended investigation of the effects arising when using the
  deprojected $\xi(r)$ instead of that directly measured (hereafter
  $\xi_{dep}$ and $\xi_{dir}$ respectively) is carried out in \citet{marulli2012}.
Here we limit the discussion to the impact of
  the deprojection technique on the estimate of $\beta$, as a function
  of the mass (i.e. the bias) of the adopted tracers, focussing on the
  systematic effects (Figure \ref{fig beta dep}).
One possible source of systematic error in performing the
de-projection is the necessity of defining a
finite integration limit $\pi_{max}$ in Eq. (\ref{eq deprojected}). 
In Figure \ref{fig beta dep} two different choices of $\pi_{max}$ are considered.
We notice that these choices (purple inverted triangles and yellow rhombs)
result in different slopes of $\beta$ as a function of bias, which
differ from the slope obtained using $\xi_{dir}$ (green triangles). 
This is plausibly due to the fact that using a limiting $\pi_{max}$ we
are underestimating the integral (consider that $\xi > 0$ for $\pi
\lesssim 100 \ h^{-1} \text{Mpc}$). This effect grows when the bias 
increases, because of the corresponding growth of $\xi$ which leads to
a larger ``loss of power'' in $w_p$. 
However, we cannot use arbitrarily large values of $\pi_{max}$ because
the  statistical error increases for larger $\pi_{max}$ (see lowest
panel of Figure \ref{fig beta dep}). This may be due to the increase
of the shot noise at large separations. 
Similarly, the drop of correlation signal at small separations due to
the finite size of the dark matter halos produces an impact on $\beta$
which grows with bias. Finally, as suggested previously
\citep{2008Natur.451..541G} and discussed 
extensively in \citet{marulli2012}, Figure \ref{fig beta dep} shows how
using $\xi_{dep}$ in modelling RSD, produces a statistical error about
twice as large as that obtained using $\xi_{dir}$ (lower panel).  

\bsp

\label{lastpage}


\begin{thebibliography}{99}

\bibitem[\protect\citeauthoryear{Acquaviva et al.}{2008}]
{Ac} Acquaviva V., Hajian A., Spergel D.~N., Das S., 2008, PhRvD, 78, 043514 

\bibitem[\protect\citeauthoryear{Alcock \& Paczynski}{1979}]
{1979Natur.281..358A} Alcock C., Paczynski B., 1979, Nature, 281, 358 

\bibitem[\protect\citeauthoryear{Angulo et al.}{2008}]
{2008MNRAS.383..755A} Angulo R.~E., Baugh C.~M., Frenk C.~S., Lacey C.~G., 2008, MNRAS, 383, 755 

\bibitem[\protect\citeauthoryear{Angulo \& White}{2010}]
{angulo_white 2010} Angulo, R.E., White, S.D.M., 2010, MNRAS, 405, 143

\bibitem[\protect\citeauthoryear{Bianchi}{2010}]{Bianchi2010} Bianchi
    D., 2010, Master Laurea Thesis, Unversity of Milan

\bibitem[\protect\citeauthoryear{Blake et al.}{2011}]
{2011MNRAS.415.2876B} Blake C., et al., 2011, MNRAS, 415, 2876 

\bibitem[\protect\citeauthoryear{Bueno Belloso, Garc{\'{\i}}a-Bellido, \& Sapone}{2011}]
{2011JCAP...10..010B} Bueno Belloso A., Garc{\'{\i}}a-Bellido J., Sapone D., 2011, JCAP, 10, 10 

\bibitem[\protect\citeauthoryear{Cabr{\'e} \& Gazta{\~n}aga}{2009}]
{2009MNRAS.393.1183C} Cabr{\'e} A., Gazta{\~n}aga E., 2009, MNRAS, 393, 1183 

\bibitem[\protect\citeauthoryear{Cappelluti et al.}{2011}]
{2011MSAIS..17..159C} Cappelluti N., et al., 2011, MSAIS, 17, 159 



\bibitem[\protect\citeauthoryear{Davis \& Peebles}{1983}]
{1983ApJ...267..465D} Davis M., Peebles P.~J.~E., 1983, ApJ, 267, 465 

\bibitem[\protect\citeauthoryear{Davis et al.}{1985}]
{1985ApJ...292..371D} Davis M., Efstathiou G., Frenk C.~S., White S.~D.~M., 1985, ApJ, 292, 371 

\bibitem[\protect\citeauthoryear{de la Torre \& Guzzo}{2012}]
{2012arXiv1202.5559D} de la Torre S., Guzzo L., 2012, arXiv, arXiv:1202.5559 


\bibitem[\protect\citeauthoryear{De Lucia \& Blaizot}{2007}]
{delucia07} De Lucia G., Blaizot J., 2007, MNRAS, 375, 2 

\bibitem[\protect\citeauthoryear{di Porto, Amendola, \& Branchini}{2012}]
{2012MNRAS.419..985D} di Porto C., Amendola L., Branchini E., 2012, MNRAS, 419, 985

\bibitem[\protect\citeauthoryear{Eisenstein et al.}{2011}]
{2011AJ....142...72E} Eisenstein D.~J., et al., 2011, AJ, 142, 72 

\bibitem[\protect\citeauthoryear{Feldman, Kaiser, \& Peacock}{1994}]
{1994ApJ...426...23F} Feldman H.~A., Kaiser N., Peacock J.~A., 1994, ApJ, 426, 23 

\bibitem[\protect\citeauthoryear{Fisher et al.}{1994}]
{1994MNRAS.266...50F} Fisher K.~B., Davis M., Strauss M.~A., Yahil A., Huchra J., 1994, MNRAS, 266, 50 

\bibitem[\protect\citeauthoryear{Fisher et al.}{1994}]
{1994MNRAS.267..927F} Fisher K.~B., Davis M., Strauss M.~A., Yahil A., Huchra J.~P., 1994, MNRAS, 267, 927  

\bibitem[\protect\citeauthoryear{Fry}{1985}]
{1985PhLB..158..211F} Fry J.~N., 1985, PhLB, 158, 211

\bibitem[\protect\citeauthoryear{Guzzo et al.}{2008}]
{2008Natur.451..541G} Guzzo L., et al., 2008, Nature, 451, 541 

\bibitem[\protect\citeauthoryear{Hamilton}{1992}]
{1992ApJ...385L...5H} Hamilton A.~J.~S., 1992, ApJ, 385, L5 

\bibitem[\protect\citeauthoryear{Hamilton}{1993}]
{1993ApJ...417...19H} Hamilton A.~J.~S., 1993, ApJ, 417, 19

\bibitem[\protect\citeauthoryear{Hamilton}{1998}]
{Ha} Hamilton, A.~J.~S. 1998,
in D. Hamilton, ed,
The Evolving Universe.
Kluwer, Dordrecht, p. 185

\bibitem[\protect\citeauthoryear{Hawken et al.}{2012}]
{hawken2012} Hawken A. J., Abdalla F. B., H{\"u}tsi G., Lahav O., 2012, MNRAS, 424, 2 

\bibitem[\protect\citeauthoryear{Hawkins et al.}{2003}]
{2003MNRAS.346...78H} Hawkins E., et al., 2003, MNRAS, 346, 78 

\bibitem[\protect\citeauthoryear{Hewett}{1982}]
{1982MNRAS.201..867H} Hewett P.~C., 1982, MNRAS, 201, 867
 
\bibitem[\protect\citeauthoryear{Jenkins et al.}{2001}]
{2001MNRAS.321..372J} Jenkins A., Frenk C.~S., White S.~D.~M., 
Colberg J.~M., Cole S., Evrard A.~E., Couchman H.~M.~P., Yoshida N., 2001, MNRAS, 321, 372 

\bibitem[\protect\citeauthoryear{Jennings, Baugh, \& Pascoli}{2011}]
{2011MNRAS.410.2081J} Jennings E., Baugh C.~M., Pascoli S., 2011, MNRAS, 410, 2081


\bibitem[\protect\citeauthoryear{Kaiser}{1987}]
{1987MNRAS.227....1K} Kaiser N., 1987, MNRAS, 227, 1 


\bibitem[\protect\citeauthoryear{Kwan, Lewis, \& Linder}{2012}]
{2011arXiv1105.1194K} Kwan J., Lewis G. F., Linder E. V., 2012, ApJ, 748, 78 

\bibitem[\protect\citeauthoryear{Landy \& Szalay}{1993}]
{1993ApJ...412...64L} Landy S.~D., Szalay A.~S., 1993, ApJ, 412, 64 

\bibitem[\protect\citeauthoryear{Larson et al.}{2011}]
{2011ApJS..192...16L} Larson D., et al., 2011, ApJS, 192, 16 


\bibitem[\protect\citeauthoryear{Laureijs et al.}{2011}]
{2011arXiv1110.3193L} Laureijs R., et al., 2011, arXiv, arXiv:1110.3193 

\bibitem[\protect\citeauthoryear{Lightman \& Schechter}{1990}]
{1990ApJS...74..831L} Lightman A.~P., Schechter P.~L., 1990, ApJS, 74, 831 

\bibitem[\protect\citeauthoryear{Linder}{2008}]
{Li2} Linder E.~V., 2008, APh, 29, 336 

\bibitem[\protect\citeauthoryear{Marulli et al.}{2012}]
{marulli2012} Marulli F., Bianchi D., Branchini E., 
Guzzo L., Moscardini L., Angulo R. E., 2012, MNRAS, 426, 2566 

\bibitem[\protect\citeauthoryear{McDonald \& Seljak}{2009}]
{2009JCAP...10..007M} McDonald P., Seljak U., 2009, JCAP, 10, 7 

\bibitem[\protect\citeauthoryear{Nesseris \& Perivolaropoulos}{2008}]
{Ne} Nesseris S., Perivolaropoulos L., 2008, PhRvD, 77, 023504 

\bibitem[\protect\citeauthoryear{Okumura \& Jing}{2011}]
{2011ApJ...726....5O} Okumura T., Jing Y.~P., 2011, ApJ, 726, 5 

\bibitem[\protect\citeauthoryear{Peacock}{1999}]
{Peacock} Peacock J. A., 1999, Cosmological Physics, Cambridge Univ. Press, Cambridge

\bibitem[\protect\citeauthoryear{Peebles}{1980}]
{1980lssu.book.....P} Peebles P.~J.~E., 1980, lssu.book,  

\bibitem[\protect\citeauthoryear{Percival \& White}{2009}]
{PW} Percival W.~J., White M., 2009, MNRAS, 393, 297 

\bibitem[\protect\citeauthoryear{Percival et al.}{2010}]
{2010MNRAS.401.2148P} Percival W.~J., et al., 2010, MNRAS, 401, 2148 

\bibitem[\protect\citeauthoryear{Perlmutter et al.}{1999}]
{Pe} Perlmutter S., et al., 1999, ApJ, 517, 565 

\bibitem[\protect\citeauthoryear{Riess et al.}{1998}]
{Ri} Riess A.~G., et al., 1998, AJ, 116, 1009 

\bibitem[\protect\citeauthoryear{Samushia et al.}{2011}]
{2011MNRAS.410.1993S} Samushia L., et al., 2011, MNRAS, 410, 1993

\bibitem[\protect\citeauthoryear{Samushia, Percival, \& Raccanelli}{2012}]
{2011arXiv1102.1014S} Samushia L., Percival W. J., Raccanelli A., 2012, MNRAS, 420, 2102

\bibitem[\protect\citeauthoryear{S{\'a}nchez et al.}{2006}]
{2006MNRAS.366..189S} S{\'a}nchez A.~G., Baugh C.~M., Percival W.~J., Peacock J.~A., Padilla N.~D., Cole S., Frenk C.~S., Norberg P., 2006, MNRAS, 366, 189  

\bibitem[\protect\citeauthoryear{Saunders, Rowan-Robinson, \& Lawrence}{1992}]
{1992MNRAS.258..134S} Saunders W., Rowan-Robinson M., Lawrence A.,
1992, MNRAS, 258, 134  

\bibitem[\protect\citeauthoryear{Scoccimarro}{2004}]
{2004PhRvD..70h3007S} Scoccimarro R., 2004, PhRvD, 70, 083007 


\bibitem[\protect\citeauthoryear{Seo \& Eisenstein}{2003}]
{2003ApJ...598..720S} Seo H.-J., Eisenstein D.~J., 2003, ApJ, 598, 720 


\bibitem[\protect\citeauthoryear{Sheth, Mo, \& Tormen}{2001}]
{2001MNRAS.323....1S} Sheth R.~K., Mo H.~J., Tormen G., 2001, MNRAS, 323, 1 

\bibitem[\protect\citeauthoryear{Simpson \& Peacock}{2010}]
{2010PhRvD..81d3512S} Simpson F., Peacock J.~A., 2010, PhRvD, 81, 043512 

\bibitem[\protect\citeauthoryear{Song \& Percival}{2009}]
{SP} Song Y.-S., Percival W.~J., 2009, JCAP, 10, 4


\bibitem[\protect\citeauthoryear{Springel et al.}{2005}]
{springel05} Springel V., et al., 2005, Natur, 435, 629 

\bibitem[\protect\citeauthoryear{Taruya, Nishimichi, \& Saito}{2010}]
{2010arXiv1006.0699T} Taruya A., Nishimichi T., Saito S., 2010, PhRvD, 82, 063522

\bibitem[\protect\citeauthoryear{Tegmark}{1997}]
{1997PhRvL..79.3806T} Tegmark M., 1997, PhRvL, 79, 3806 

\bibitem[\protect\citeauthoryear{Tinker et al.}{2010}]
{2010ApJ...724..878T} Tinker J.~L., Robertson B.~E., Kravtsov A.~V., Klypin A., Warren M.~S., Yepes G., Gottl{\"o}ber S., 2010, ApJ, 724, 878 

\bibitem[\protect\citeauthoryear{Tinker, Weinberg, \& Zheng}{2006}]
{2006MNRAS.368...85T} Tinker J.~L., Weinberg D.~H., Zheng Z., 2006, MNRAS, 368, 85 


\bibitem[\protect\citeauthoryear{Wang \& Steinhardt}{1998}]
{1998ApJ...508..483W} Wang L., Steinhardt P.~J., 1998, ApJ, 508, 483 

\bibitem[\protect\citeauthoryear{Wang}{2008}]
{Wn} Wang Y., 2008, JCAP, 5, 21 

\bibitem[\protect\citeauthoryear{Wang et al.}{2010}]
{2010MNRAS.409..737W} Wang Y., et al., 2010, MNRAS, 409, 737


\bibitem[\protect\citeauthoryear{White, Song, \& Percival}{2009}]
{2009MNRAS.397.1348W} White M., Song Y.-S., Percival W.~J., 2009, MNRAS, 397, 1348 

\bibitem[\protect\citeauthoryear{Zhang et al.}{2007}]
{ZhangNEW} Zhang, P., Liguori, M., Bean, R., \& Dodelson, S. 2007, Physical Review Letters, 99, 141302

\bibitem[\protect\citeauthoryear{Zurek et al.}{1994}]
{1994ApJ...431..559Z} Zurek W.~H., Quinn P.~J., Salmon J.~K., Warren M.~S., 1994, ApJ, 431, 559

\end{thebibliography}
\end{document}